\documentclass{article}
\usepackage[numbers]{natbib}
\usepackage{authblk}
\usepackage{url}
\usepackage{amssymb}
\usepackage{amsmath}
\usepackage{bm}
\usepackage{relsize}
\usepackage{graphicx}
\begin{document}
	
\title{Neutron Skins: Weak Elastic Scattering and Neutron Stars }
\author[1]{Juliette M. Mammei}
\author[2]{Charles J. Horowitz}
\author[3]{Jorge Piekarewicz}
\author[4]{Brendan Reed}
\author[5]{Concettina Sfienti}
\affil[1]{$^1$Department of Physics and Astronomy, University of Manitoba, Winnipeg, MB, Canada, R3T4E7; email: jmammei@physics.umanitoba.ca}
\affil[2]{$^2$Department of Physics, Indiana University, Bloomington, IN 47405, USA; email: horowit@indiana.edu}
\affil[3]{$^3$Department of Physics, Florida State University
		Tallahassee, FL 32306-4350}
\affil[4]{$^4$Theoretical Division, Los Alamos National Laboratory, Los Alamos, New Mexico 87545, USA}
\affil[5]{$^5$Institut für Kernphysik, Johannes Gutenberg-Universität Mainz, Germany, D-55099}


\maketitle



\begin{abstract}
The recently completed PREX-2 campaign---which measured the weak form factor of lead at an optimal momentum transfer---has confirmed that 
the neutron skin of lead is relatively large and has provided a precise determination of the interior baryon density of a heavy nucleus. In turn, the
measured form factor can be related to various nuclear and neutron-star properties. Astrophysical observations by the NICER mission have benefited
from improvements in flux, energy resolution, and notably, timing resolution. NICER has the capability to measure pulse profile data, which enables 
simultaneous mass-radius determinations. During the next decade, measurements in astrophysics, gravitational-wave astronomy, and nuclear physics 
are expected to provide a wealth of more precise data. In this review we provide an overview of the current state of neutron skin measurements and 
offer insights into the prospects for the future.
\end{abstract}

\maketitle

\section{INTRODUCTION} 
We are living in a very exciting time for nuclear physics and astrophysics. The ability to measure properties of individual nucleons and complex 
nuclei with ever higher precision will inform nuclear theory in areas of relevance to the extreme environments found within neutron stars. In particular, 
parity-violating electron scattering (PVES) is a powerful, yet challenging, experimental method that has been employed in experiments like PREX-2 
and CREX. By relying exclusively on the electroweak interaction, PVES uncovers the inner workings of atomic nuclei in a model independent way. 
Unlike hadronic experiments that suffer from large and uncontrolled systematic errors, PVES offers sharp and critical insights into fundamental aspects 
of the nuclear dynamics. Furthermore, the scientific community is benefiting greatly from the emergence of neutrino, electromagnetic and gravitational-wave 
observatories worldwide, spearheading the new era of multi-messenger astronomy. The deployment of advanced telescopes such as the 
Neutron-star Interior Composition ExploreR (NICER), the Chandra X-ray Observatory, and the XMM-Newton mission among others, is revolutionizing 
our ability to observe and analyze neutron stars. In turn, the historic detection of gravitational waves from binary neutron-star mergers is providing 
fundamental new insights into the astrophysical site for the creation of the heavy elements and on the nature of dense matter. Together with future PVES 
experiments, these telescopes will provide fundamental information on the nuclear equation of state (EOS), which quantifies how the pressure support against
gravitational collapse evolves as a function of density.  In this article, we will describe the profound connection between neutron stars and nuclei.  We will 
discuss neutron rich matter and the observables that are particularly sensitive to the equation of state. We will describe the PREX-2 and CREX campaigns
and the extraction of model independent weak form factors. Finally, we will survey some future measurements---both earth- and space-based---which
will further our understanding of the nuclear theory needed to understand the structure, dynamics, and composition of neutron stars.

\section{Neutron stars and neutron rich matter EOS} 
Neutron stars are compact objects more massive than the Sun but only about 12 km in radius\,\cite{doi:10.1126/science.1090720,doi:10.1146/annurev-nucl-102419-124827}. 
The first neutron star (NS) was discovered as a radio pulsar in 1967 by Jocelyn Bell-Burnell \cite{JBell1968}. The densest known objects this side of black holes, their central density 
is at least a few times larger than the density found in the interior of atomic nuclei, providing exciting laboratories to study quantum chromodynamics (QCD) as well as 
gravitational and electromagnetic fields under extreme conditions.  In this section we summarize the most salient features of neutron rich matter.  

\subsection{Neutron star structure}

\begin{figure}[tb]
\includegraphics[width=5.0in]{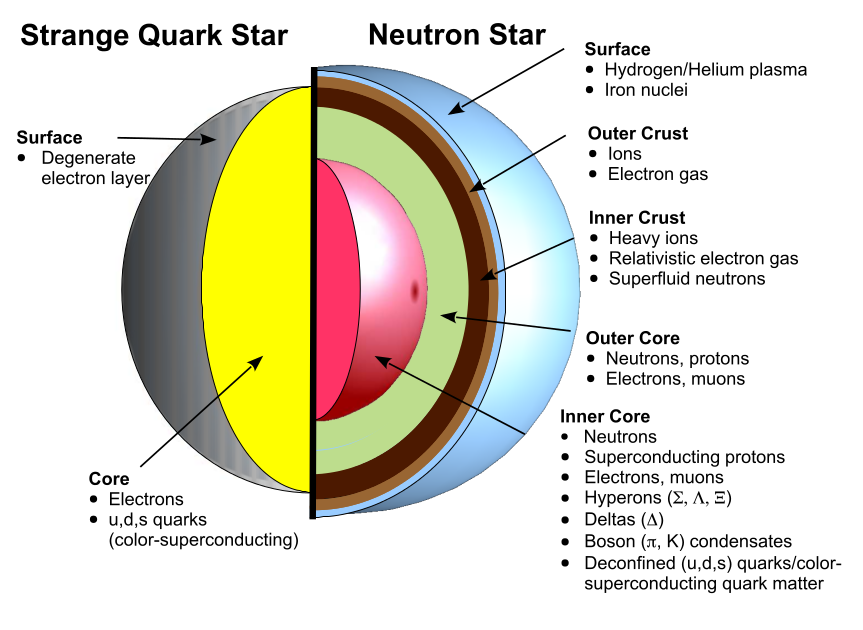}
\caption{The assumed structure of a neutron star\cite{peeling}, consisting of a surface, an outer and inner crust, and an outer and inner core. These regions are defined by dramatic changes in structure and composition. The traditional view of a neutron star is compared against a possible model of a strange quark star, which because of a presumed phase transition, often result in a smaller radius for a given mass.}
\label{fig:quark_NStar}
\end{figure}

The structure of a neutron star is described theoretically by the Tolman-Oppenheimer-Volkoff (TOV) equations\,\cite{PhysRev.55.374}, which represent the generalization
of Newtonian gravity to the domain of general relativity. The structure of a neutron star is not completely understood--even the maximum mass of a star is not known---but 
the general description (see Fig. \ref{fig:quark_NStar}) is that there is an atmosphere and an envelope---which encompass the stellar surface---an outer and inner crust, and an outer and inner core.  
The crust is composed of a crystal lattice of neutron rich nuclei and relativistic electrons, while the outer core is a quantum liquid of neutron rich matter.  The TOV equations 
describe the hydrostatic equilibrium where the force of gravity is balanced by the pressure of neutron rich matter.  The mass and radius of a neutron star depend on the 
interactions in the dense matter of the star.

The composition of the inner core, in particular, is a mystery.  Besides neutrons, protons, electrons, and muons, the inner core may contain exotic degrees of freedom such as hyperons (baryons containing strange quarks), meson (or boson) condensates, and deconfined quark matter. The emergence of new degrees of freedom is accompanied by a softening of the EOS, which often results in both smaller radii and a smaller maximum mass. In the case of models that predict the emergence of hyperons, this problem is commonly referred to in the literature as the ``hyperon puzzle''\,\cite{Vidana:2000mx,Oertel:2016xsn,Oertel:2016bki}, since the predicted maximum neutron-star mass of some hyperon models is inconsistent with the observation of $2M_{\odot}$ neutron stars\ \cite{Demorest:2010bx,Antoniadis:2013pzd,
Cromartie:2019kug,Fonseca:2021wxt}.

Also possible is the existence of strange quark stars, or at least stars with high-density cores made of equal numbers of 
massless up, down, and strange quarks. In this limit, the ground state is a color superconductor with a unique pairing 
scheme that couples color and flavor\,\cite{Alford:2007xm}. Unfortunately, it is now believed that the extreme densities required for such a phase to emerge can not be reached in the stellar cores. So assessing the impact of QCD at the densities of relevance to neutron stars remains an important challenge.

\subsection{Neutron rich matter equation of state}

\begin{figure}[tb]
\includegraphics[width=3.5in]{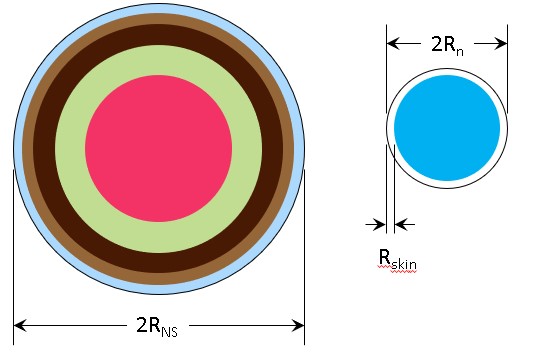}
\caption{A neutron star (left) is 18 orders of magnitude larger than a Pb nucleus (right, not to scale).  Nevertheless, the star is made of the same neutrons with the same strong interactions and equation of state.  The structure of both objects depends on the pressure of neutron rich matter $P$.  This pressure pushes neutrons out against surface tension and increases the neutron radius $R_n$ of $^{208}$Pb.  The radius of a neutron star $R_{NS}$ also depends on $P$.  Therefore a
measurement of $R_n$ in the laboratory has important implications for the structure of neutron stars.}
\label{fig:NS_Pb}
\end{figure}

\begin{figure}[h]
\includegraphics[width=4.5in]{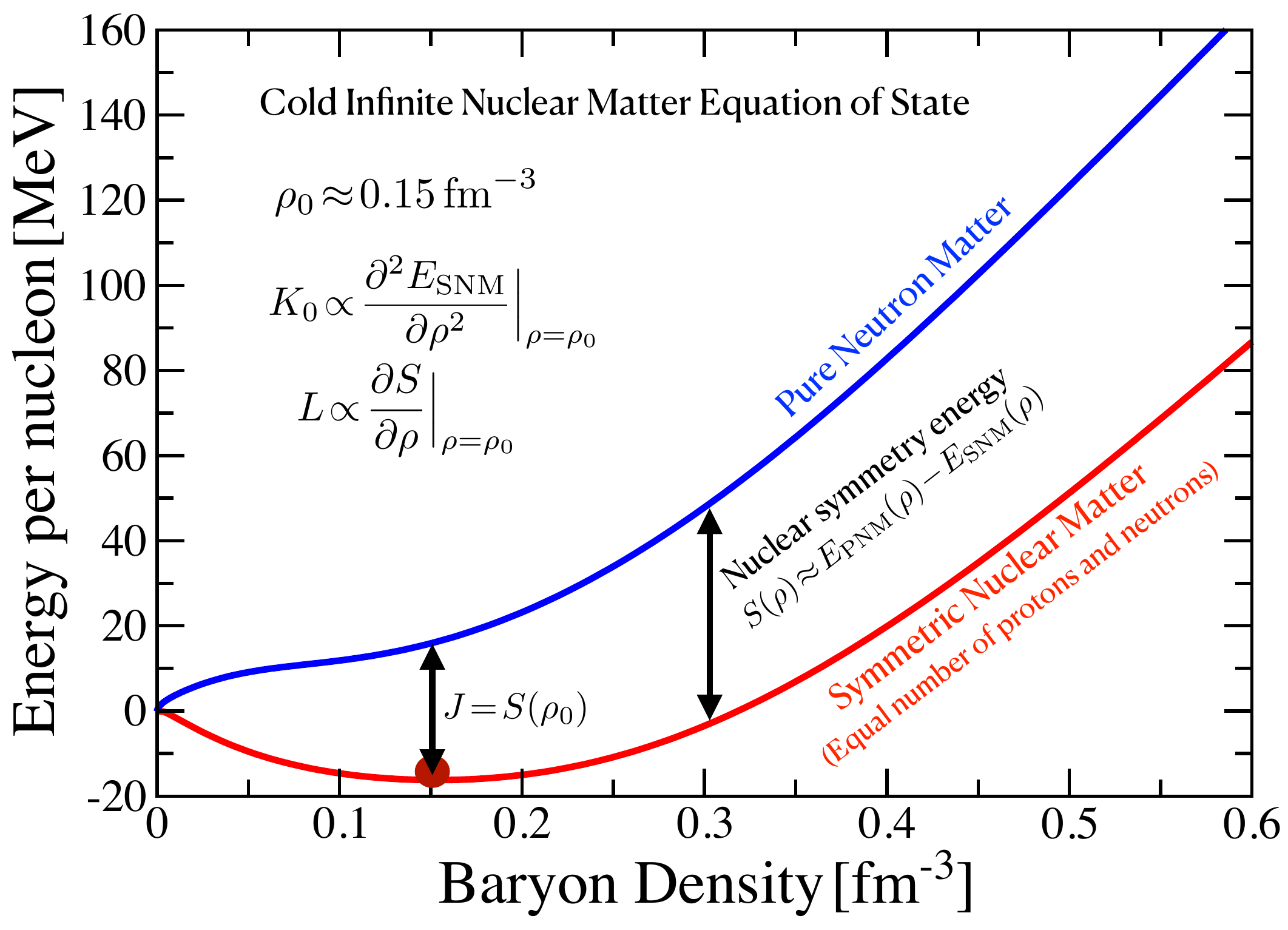}
\caption{A depiction of the nuclear equation of state. The minimum in the energy per particle of symmetric nuclear matter (red) occurs at the saturation density of $\rho_{0}\!\approx\!0.15\,{\rm fm}^{-3}$. The nuclear symmetry energy $S(\rho)$ is approximated as the difference between the energy of pure neutron matter (blue) and symmetric nuclear matter. At saturation density this value is commonly referred to as $J$. The slope of the symmetry energy $L$ is highly correlated to the neutron skin of neutron-rich nuclei as well as to the radii of neutron stars. The nuclear 
incompressibility $K_{0}$ of symmetric nuclear matter is related to the curvature at saturation density and correlates strongly to the centroid energy of giant monopole resonances.}
\label{fig:eos}
\end{figure}

The equation of state (EOS) of neutron rich matter is a relationship giving the pressure $P=P(\epsilon)$ as a function of the energy density $\epsilon$.  
For two decades the neutron skin thickness of ${}^{208}$Pb has been identified as an ideal 
laboratory observable to constrain the EOS of neutron rich matter, see Fig. \ref{fig:NS_Pb}, particularly the poorly 
determined density dependence of the symmetry energy\,\cite{Brown:2000,Furnstahl:2001un,
Centelles:2008vu,RocaMaza:2011pm}. The EOS of infinite nuclear matter at zero temperature is enshrined in the energy per particle which depends on both the conserved neutron ($\rho_{n}$) 
and proton ($\rho_{p}$) densities; here we assume that the electroweak sector has been ``turned off\,''. 
Moreover, it is customary to separate the EOS into two contributions, one that represents the energy 
of  symmetric ($\rho_{n}\!=\!\rho_{p}$) nuclear matter and another one that accounts for the breaking 
of the symmetry (see Fig. \ref{fig:eos}). That is,
\begin{equation}
  \frac{E}{A}(\rho,\alpha) -\!M \equiv {\cal E}(\rho,\alpha)
    = {\cal E}_{\rm SNM}(\rho)
    +\alpha^{2}{\cal S}(\rho)  
     + {\cal O}(\alpha^{4}) \,,
 \label{EOS}
\end {equation} 
where $\rho\!=\!(\rho_{n}\!+\!\rho_{p})$ is the total baryon density given by the sum of neutron and
proton densities, and $\alpha\!=(\rho_{n}\!-\!\rho_{p})/\rho$ is the neutron-proton asymmetry. The 
first-order correction to the energy of symmetric nuclear matter ${\cal E}_{\rm SNM}(\rho)$ is 
encoded in the symmetry energy ${\cal S}(\rho) $. The symmetry energy quantifies the increase 
in the energy per particle of infinite nuclear matter for systems with an isospin imbalance (e.g., more 
neutrons than protons). Furthermore, given the preeminent role of nuclear saturation, the energy of 
symmetric nuclear matter and the symmetry energy may be described in terms of a few bulk 
parameters that characterize their behavior around saturation density. In this review we focus on 
the density dependence of the symmetry energy\,\cite{Piekarewicz:2008nh}:
\begin{equation}
 {\cal S}(\rho) = J + L \frac{(\rho-\rho_0)}{3\rho_0} + \ldots   
\label{EandS}
\end{equation}
The first term ($J$) represents the correction to the binding energy of symmetric nuclear matter,
whereas the second term ($L$) dictates how rapidly the symmetry energy increases with
density. It is the slope of the symmetry energy $L$ that displays a strong correlation to the neutron 
skin thickness of ${}^{208}$Pb. Given that symmetric nuclear matter saturates, namely, its pressure 
vanishes at saturation, the slope of the symmetry energy $L$ is closely related to the pressure of 
pure neutron matter at saturation density. That is,  
\begin{equation}
 P_{\,\rm PNM}(\rho_0)\!\approx\!\frac{1}{3} L\rho_0.  
\label{Ppnm}
\end{equation}

\subsection{X-ray measurements of NS radii}

Observations of NS masses and radii can help determine the EOS.  Many NS have surface temperatures near $10^6$ K and emit X-rays.  Recently the NICER X-ray telescope has inferred masses and radii of some X-ray pulsars \cite{Miller:2019cac,Riley:2019yda,Miller:2021qha}.  These measurements depend on pulse profile modeling of the light curve as a neutron star with hot spots rotates.  When a hot spot is on the far side of the star some fraction of the radiation is still visible because of the gravitational curvature of space.   Thus the depth of the light curve minimum provides a measure of the curvature and this depends on the star's mass and radius.  We discuss this further in Sec.~\ref{Subsection.Nicerfuture}.

\subsection{Gravitational wave measurements of NS deformability}

The historic detection of gravitational waves emitted from the binary neutron star merger GW170817 
has opened a brand new window into the Universe\,\cite{Abbott:PRL2017}. In particular, GW170817 is 
providing new insights into the astrophysical site for the rapid-neutron capture process - believed to be the mechanism behind the production of about half of the heavy elements beyond
iron\,\cite{Drout:2017ijr,Cowperthwaite:2017dyu,Chornock:2017sdf,Nicholl:2017ahq} -and on the structure, 
dynamics, and composition of neutron stars\,\cite{Bauswein:2017vtn, Fattoyev:2017jql,Annala:2017llu, 
Abbott:2018exr,Most:2018hfd,Tews:2018chv,Malik:2018zcf,Radice:2017lry, Radice:2018ozg, Tews:2019cap,
Capano:2019eae,Tsang:2019mlz,Tsang:2020lmb,Drischler:2020hwi,Landry:2020vaw,Xie:2020rwg,Essick:2021kjb,
Chatziioannou:2021tdi,Reed:2021nqk,Sammarruca:2022ser}

Binary neutron star mergers provide a few critical observables that inform the EOS of neutron rich matter. First, the chirp mass, defined as a linear combination of the individual masses of the two stars, 
\begin{equation}
 {\mathcal {M}}={\frac {(M_{1}M_{2})^{3/5}}{(M_{1}+M_{2})^{1/5}}},
\end{equation}
is the most precise observable extracted from the gravitation wave profile\,\cite{Abbott:PRL2017}. 
Second, the tidal deformability encodes how a neutron star deforms in response to the tidal field generated by its companion star\,\cite{Damour:1991yw,Flanagan:2007ix}. In the linear regime, the
constant of proportionality connecting the external tidal forces to the star's mass quadrupole is the
dimensionless tidal deformability $\Lambda$ defined as
\begin{equation}
 \Lambda = \frac{2}{3}k_{2}\left(\frac{c^{2}R}{GM}\right)^{5}
 \label{Lambda}
\end{equation}
where $k_{2}$ is the second Love number\,\cite{Love:1909,Binnington:2009bb,Damour:2012yf}.
Although $k_{2}$ is sensitive to the underlying EOS, most of the sensitivity is contained
in the compactness parameter $(c^{2}R)/(GM)$\,\cite{Hinderer:2007mb,Hinderer:2009ca,
Damour:2009vw,Postnikov:2010yn,Fattoyev:2012uu,Steiner:2014pda,Fattoyev:2017jql,
Piekarewicz:2018sgy}. An estimate by the LIGO-Virgo collaboration of the tidal deformability of a 
1.4\,$M_{\rm sun}$ neutron star yields the relatively small value of $\Lambda_{1.4}\!\lesssim\!580$,
suggesting that neutron stars are dense objects with a small radius that are difficult to 
deform\,\cite{Abbott:2018exr}. This implies a fairly soft EOS at 
intermediate densities. To our knowledge, GW170817 provides one of the very few indications that 
the EOS is soft.
\def\Cevens{CEvNS\hspace{3pt}}
\def\expo{\mathlarger{\mathlarger{e}}}
\def\alphad{$\!\alpha_{\raisebox{-1pt}{\tiny  D}}\,$}
\def\fF{\mathlarger{f}_{\raisebox{-0.5pt}{\!\tiny F}}}
\def\fS{\mathlarger{f}_{\raisebox{-0.5pt}{\!\tiny SF}}}
\newcommand{\FF}[1]{{F}_{\raisebox{-0.5pt}{\!\tiny #1}}}
\newcommand{\rhoX}[1]{{\rho}_{\raisebox{-3.50pt}{\!\tiny #1}}}

\section{Nuclear physics observables related to neutron rich matter}
Given the importance of neutron skins in constraining the equation of state, in this section we describe our knowledge
of nuclear sizes from both parity-conserving and parity-violating elastic electron scattering---highlighting two recent 
experimental efforts. While the main focus of this review is parity-violating electron scattering, we highlight its connection 
to coherent elastic neutrino nucleus scattering which also offers a promising model independent method for determining 
neutron densities.

\subsection{Nuclear sizes and elastic electron-nucleus scattering}
\label{sec:ENS}

The distribution of electric charge is a fundamental nuclear property that has been 
mapped with great accuracy along the valley of stability. Elastic electron scattering 
experiments pioneered by Hofstadter in the late 1950's\,\cite{Hofstadter:1956qs} 
and that continue to improve to this day\,\cite{DeJager:1987qc,Fricke:1995,Angeli:2013,
Suda:2009zz,Suda:2017nss}, 
have painted the most compelling picture of the spatial charge distribution. 
Given that the electric charge of the nucleus is carried by the protons, elastic electron 
scattering provides a powerful tool for the determination of the ground-state proton density.

To appreciate the power of electron scattering in the determination of the charge
distribution of atomic nuclei, we provide an expression for the cross section in the
laboratory frame for when the electron scatters elastically from a spinless target. 
That is\,\cite{Aitchinson:1982},
\begin{equation}
    \left(\frac{d\sigma}{d\Omega}\right) = 
    \left[\frac{\alpha^{2}\cos^{2}(\theta/2)}{4E^{2}\sin^{4}(\theta/2)}\!\!\left(\frac{E'}{E}\right)\right]
    Z^{2}F_{\rm ch}^{2}(Q^{2}),
\label{dsdOmega}
\end{equation}
where $Z$ is the electric charge of the nucleus, $\alpha$ is the fine-structure constant, $\theta$
is the scattering angle, and $E$ and $E'$ are the energies of the incoming and scattered electrons,
respectively. The expression in brackets is the Mott cross section which represents the scattering of 
a relativistic electron from a spinless and structureless target. 
 
Deviations from the structureless limit are encoded in the charge form factor of the nucleus $F_{\rm ch}(Q^{2})$ 
which depends on the exchanged photon four-momentum square 
$Q^{2}\!\equiv\!{\bf q}^{2}\!-\!\omega^{2}\!\approx\!{\bf q}^{2}$. 
 
Given that the form factor may be viewed as the Fourier transform of the spatial distribution, elastic 
electron scattering paints the most accurate picture of the distribution of charge in atomic 
nuclei\,\cite{Walecka:2001}.

To further illustrate the power of elastic electron-nucleus scattering in accurately determining nuclear sizes, 
we offer insights derived from the ``symmetrized Fermi function'', which is defined as 
follows\,\cite{Sprung:1997, Piekarewicz:2016vbn}:
\begin{equation}
 {\large\rhoX{SF}}(r) = {\large{\rhoX{0}}}\,\frac{\sinh(c/a)}{\cosh{(r/a)}+\cosh(c/a)}, 
 \hspace{3pt}\text{where}\hspace{5pt}  {\large\rhoX{0}}  = \frac{3Z}{4\pi c\left(c^{2}+\pi^{2}a^{2}\right)}.
 \label{RhoSF} 
\end{equation}
where $c$ is the half-density radius and $a$ the surface diffuseness. Although practically identical to the 
well-known two-parameter Fermi function, the symmetrized version displays better analytical properties that 
enables its Fourier transform---namely, the associated form factor---to be evaluated in closed form\,\cite{Sprung:1997}. 
That is,
\begin{equation}
 \FF{SF}(q) = \frac{3}{qc\Big((qc)^{2}+(\pi qa)^{2}\Big)}       
                  \left(\frac{\pi qa}{\sinh(\pi qa)}\right)
                  \left[\frac{\pi qa}{\tanh(\pi qa)}\sin(qc)-qc\cos(qc)\right]\,,
 \label{FFSF}                  
\end{equation}
Particularly insightful is the behavior of the symmetrized Fermi form factor in the limit of large momentum 
transfers, namely\,\cite{Piekarewicz:2016vbn}:
\begin{equation}
 \FF{SF}(q) \rightarrow
 -6\frac{\pi a}{\sqrt{c^{2}+\pi^{2}a^{2}}}\frac{\cos(qc+\delta)}{qc}\,
 \mathlarger{e^{-\pi aq}}\,; \quad 
 \tan\delta\!\equiv\!\frac{\pi a}{c}\,.
  \label{HighqFs}
\end{equation}
This expression encapsulates many of the insights developed in the context of the conventional Fermi function,
namely, diffractive oscillations controlled by the half-density radius $c$ and an exponential falloff driven by 
the diffuseness parameter $a$, or rather $\pi a$\,\cite{Amado:1979st,Amado:1986pm}. Yet, unlike the conventional 
Fermi function, the results presented here for the symmetrized version are exact. As such, all moments of the 
spatial distribution may also be evaluated exactly. For example, the mean square radius is given by

\begin{equation}
 R^{2} \equiv \langle r^{2} \rangle = \frac{3}{5}c^{2} + \frac{7}{5}(\pi a)^{2}.
 \label{SFMoments}
\end{equation}

\subsection{Parity violating elastic electron-nucleus scattering}
\label{sec:PVES}

Given that photons are insensitive to the neutron distribution, many experimental facilities have been 
commissioned with the primary goal of mapping the neutron distribution of atomic nuclei. Most of these 
experimental facilities rely on hadronic probes to map the neutron distribution. Although high statistics is 
the hallmark of hadronic experiments, the cost for such high efficiency are large systematic uncertainties 
associated with model dependencies and uncontrolled theoretical approximations. For a recent review on 
the large suite of experimental techniques devoted to map the neutron distribution and their associated 
uncertainties see\,\cite{Thiel:2019tkm}.

To eliminate the dependence on hadronic probes, Donnelly, Dubach, and Sick realized more than three 
decades ago that parity-violating electron scattering (PVES) offers a uniquely clean probe of neutron 
densities that is free from strong-interaction uncertainties\,\cite{Donnelly:1989qs}. The pioneering Lead 
Radius EXperiment (PREX) at the Jefferson Laboratory (JLab) has fulfilled this vision by providing the 
first model-independent determination of the weak-charge form factor of ${}^{208}$Pb, albeit at a single 
value of the momentum transfer\,\cite{Abrahamyan:2012gp,Horowitz:2012tj}. 
 
Defined as the fractional difference between the elastic cross section for right-handed relative to 
left-handed polarized electrons, the parity violating asymmetry is defined as follows:
 \begin{equation}
  A_{PV}(Q^{2})= \frac{\displaystyle{\left(\frac{d\sigma}{d\Omega}\right)_{\!\!R}  - 
                                            \left(\frac{d\sigma}{d\Omega}\right)_{\!\!L}}}
                           {\displaystyle{\left(\frac{d\sigma}{d\Omega}\right)_{\!\!R}  + 
                           \left(\frac{d\sigma}{d\Omega}\right)_{\!\!L}}} =
                           \frac{G_{\!F}Q^{2}}{4\pi\alpha\sqrt{2}}
                           \frac{Q_{\rm wk}F_{\rm wk}(Q^{2})}{ZF_{ch}(Q^{2})},
\label{APV}
\end{equation}
where $G_{F}$ is the Fermi constant, $Q_{\rm wk}\!=\!NQ_{\rm wk}^{\,n}\!+\!ZQ_{\rm wk}^{\,p}$ is the weak 
charge of the nucleus, and $F_{\rm wk}(Q^{2})$ is the associated weak charge form factor, with 
$F_{\rm wk}(Q^{2}\!=\!0)\!=\!1$. Among the many advantages of measuring the parity violating asymmetry is 
that $F_{\rm wk}$ is highly sensitive to the neutron distribution. This fact follows from the small weak charge 
of the proton that is strongly suppressed by the weak mixing angle:
$Q_{\rm wk}^{\,p}\!=\!1\!-\!4\sin^{2}\!\theta_{\rm W}\!=\!0.0719\pm0.0045$\,\cite{Androic:2018kni}; instead, the 
weak charge of the neutron is large.  

Given that the charge form factor is well known\,\cite{DeJager:1987qc,Fricke:1995}, the parity violating 
asymmetry provides a model-independent determination of the weak form factor. One should note that 
Eq.\,(\ref{APV}) is valid in the plane-wave approximation. Indeed, both the parity conserving cross 
section given in Eq.(\ref{dsdOmega}) as well as the parity violating asymmetry are modified by Coulomb 
distortions. However, since Coulomb distortions are well 
understood\,\cite{Horowitz:1998vv,RocaMaza:2008cg,RocaMaza:2011pm,Koshchii:2020qkr}, we focus 
on the plane-wave versions as they provide a more intuitive picture of the underlying physics. 

\begin{figure}[h]
\includegraphics[width=4in]{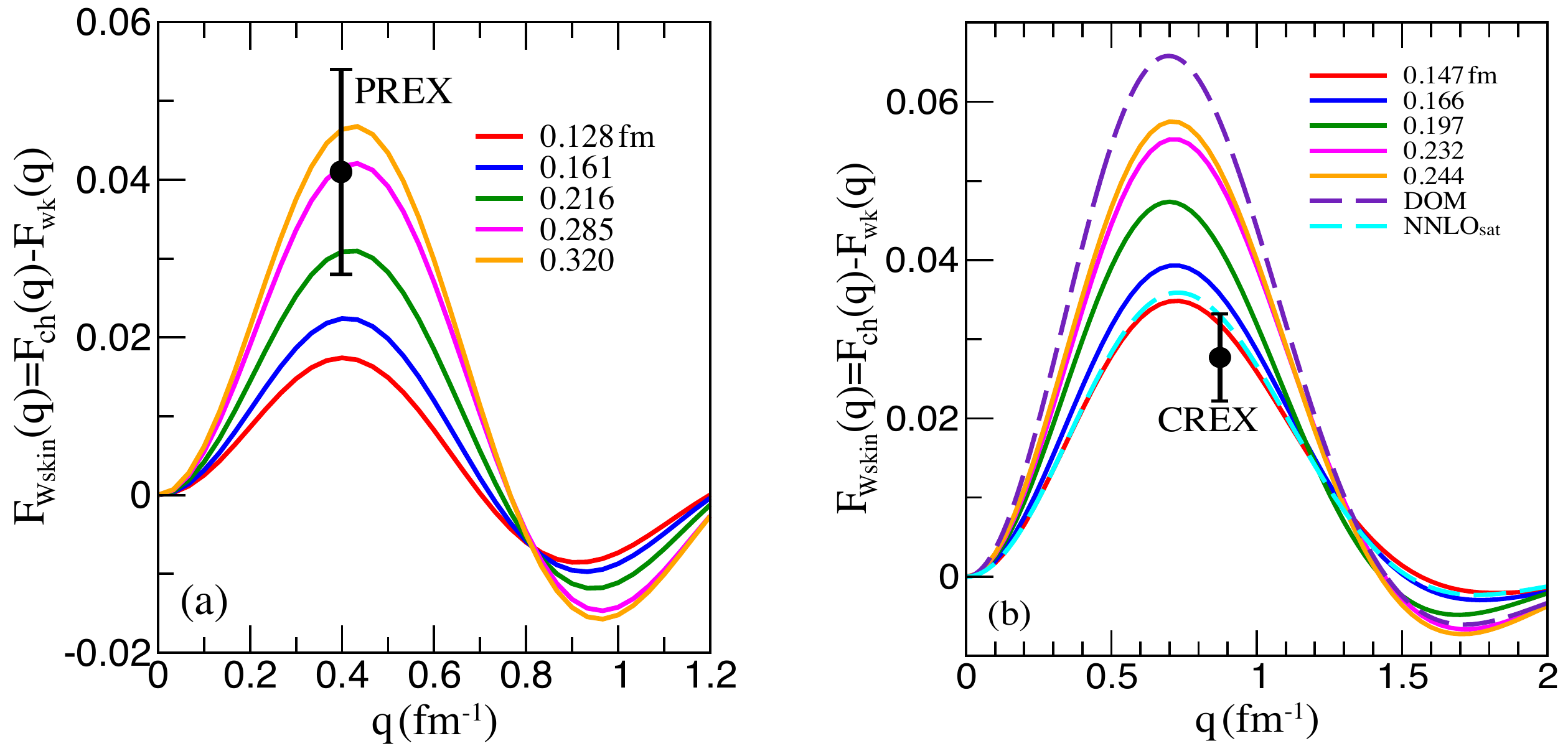}
\caption{Theoretical predictions for the weak skin form factor of (a) ${}^{208}$Pb and (b) ${}^{48}$Ca 
	       are compared against the experimental values at the single value of the momentum transfer
	       used for PREX ($0.3977\,{\rm fm}^{-1}$) and CREX ($0.8733\,{\rm fm}^{-1}$), respectively. 
	       Also shown are theoretical predictions using an ab initio approach with a chiral interaction 
	       (NNLOsat)\,\cite{Hagen:2015yea} and a dispersive optical model (DOM) 
	       approach\,\cite{Mahzoon:2017fsg}.}
\label{fig:WkSkin}
\end{figure}

The Calcium Radius Experiment (CREX)\,\cite{Adhikari:2022kgg} followed on the footsteps of the successful Lead 
Radius Experiment (PREX)\,\cite{Abrahamyan:2012gp,Horowitz:2012tj,Adhikari:2021phr}. Highlighting both experimental
campaigns is the determination of the neutron skin thickness as an observable highly sensitive to the slope of 
the symmetry energy. Particularly interesting is the fact that the slope of the symmetry energy also correlates 
strongly to the radius of low mass neutron stars\,\cite{Carriere:2002bx}.  However, given that inferring the 
neutron skin thickness from the parity violating asymmetry involves some mild model dependence, it is preferable 
to focus on the model independent form factor. 

We show in Fig.\,\ref{fig:WkSkin} predictions for the weak skin form factor, defined as the difference between 
the charge and weak form factors: $F_{\rm Wskin}(q)\!\equiv\!F_{\rm ch}(q)\!-\!F_{\rm wk}(q)$. Also shown are the 
experimental values for both ${}^{208}$Pb and ${}^{48}$Ca as quoted in Ref.\cite{Adhikari:2022kgg}. Note that by
expanding $F_{\rm Wskin}(q)$ at small momentum transfers one obtains

\begin{equation}
 F_{\rm Wskin}(q) \approx \frac{q^{2}}{6}\Big(R_{\rm wk}^{2}\!-\!R_{\rm ch}^{2}\Big). 
 \label{FWskin}
\end{equation}	   

Thus, the curvature at the origin is proportional to the $R_{\rm Wskin}\!=\!R_{\rm wk}\!-\!R_{\rm ch}$, a 
physical observable that is closely related to the neutron skin thickness 
$R_{\rm skin}\!=\!R_{n}\!-\!R_{p}$\,\cite{Horowitz:2012we}.

The large value of $F^{208}_{\rm Wskin}\!=\!0.041\pm0.013$ for ${}^{208}$Pb at the momentum transfer of the 
experiment ($0.3977{\rm fm}^{-1}$) indicates that the weak form factor falls significantly faster than the 
corresponding charge form factor, a behavior often attributed to a large value for the neutron skin. Indeed, 
the PREX collaboration reported a neutron skin thickness of
$R^{208}_{\rm skin}\!=\!0.283\pm0.071\,{\rm fm}$\,\cite{Adhikari:2021phr}. Based on these results---and 
taking advantage of the strong correlation between the neutron skin thickness of ${}^{208}$Pb and the slope 
of the symmetry energy\,\cite{Brown:2000,Furnstahl:2001un,Centelles:2008vu,RocaMaza:2011pm}---a large 
value of $L\!=\!106\pm37\,{\rm MeV}$ was deduced\,\cite{Reed:2021nqk}, which overestimates values
extracted from both theoretical approaches and experimental measurements. From such a perspective, the three
stiffest models, namely, the ones that predict the largest neutron skins in Fig.\,\ref{fig:WkSkin}(a), are all 
consistent with the 1$\sigma$ PREX result.

Given that within the context of density functional theory a robust correlation emerges between the neutron skin 
thickness of medium to heavy neutron rich nuclei and the corresponding neutron skin thickness of ${}^{208}$Pb
(see Refs.\,\cite{Piekarewicz:2012pp,Piekarewicz:2021jte} and references contained therein) it came as a surprise 
to many to learn that CREX reported an unexpectedly small value for the weak skin form factor of ${}^{48}$Ca of
$F^{48}_{\rm Wskin}\!=\!0.0277\pm0.0055$ suggesting, in turn, a very thin neutron skin of 
$R^{48}_{\rm skin}\!=\!0.121\pm0.035\,{\rm fm}$\,\cite{Adhikari:2022kgg}. In this case---and in sharp contrast to
the PREX analysis---only the softest of the five models is consistent with the CREX value. Besides the predictions 
from the five covariant EDFs, we also display in Fig.\,\ref{fig:WkSkin}(b) theoretical predictions from an ab initio 
approach\,\cite{Hagen:2015yea} and from a dispersive optical model (DOM) framework\,\cite{Mahzoon:2017fsg}. 
We note that the success of the ab initio prediction is noteworthy, especially as one of the main motivations behind 
CREX was the use of ab initio models to inform and improve the isovector sector of energy density functionals.

\begin{figure}[h]
\includegraphics[width=5in]{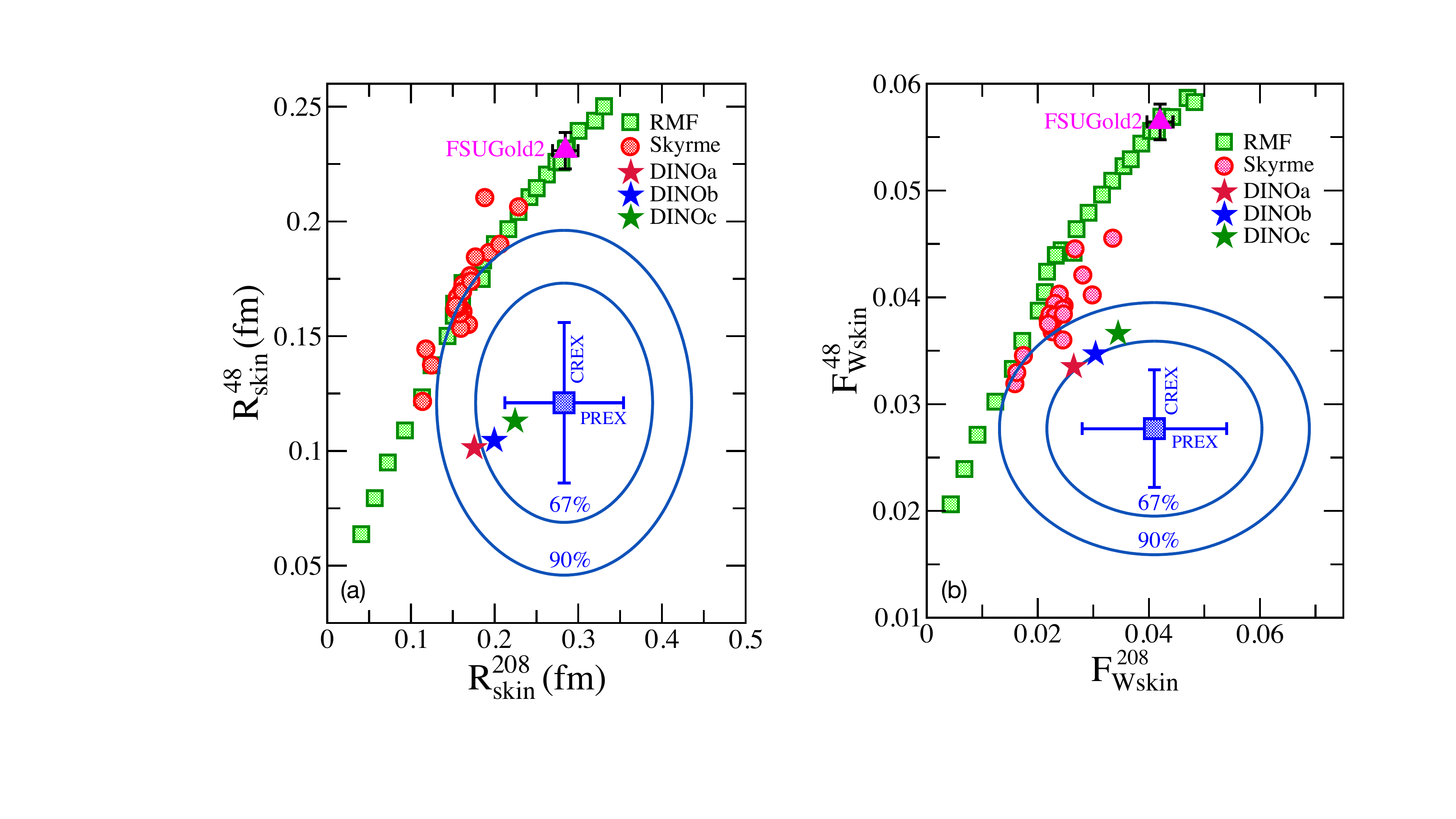}
\caption{Predictions for the (a) the neutron skin thickness and (b) weak skin form factor of ${}^{48}$Ca and ${}^{208}$Pb for a large set of covariant (green squares) and a limited set of non-relativistic (pink) EDFs.  The blue ellipses represent joint PREX and CREX 67\% and 90\% probability contours. The FSUGold2 prediction 
is included to illustrate typical statistical uncertainties. Finally, depicted with stars are the prediction of the three ``DINO" models\,\cite{Reed:2023cap}.}
\label{fig:WkSkinCompare}
\end{figure}

To further illustrate the tension in reconciling the PREX and CREX results, we display in Fig.\,\ref{fig:WkSkinCompare} 
predictions for the neutron skin thickness and weak skin form factor of ${}^{48}$Ca and ${}^{208}$Pb for a large collection
of covariant (RMF) and a limited set non relativistic (Skyrme) EDFs\,\cite{Reed:2023cap}. 

Also shown are experimental 1$\sigma$ error bars together 
with 67\% and 90\% probability contours. As it is clearly visible in the figure, none of these large number of EDFs can 
simultaneously describe the PREX and CREX results. Thus, in an effort to accommodate both results and motivated 
by a novel correlations obeyed by the covariant EDFs, a new set of ``DINO" functionals were calibrated using as input 
binding energies and charge radii of spherical nuclei as well as the PREX and CREX constraints. These results, 
depicted by the three stars, sit at or within of the 67\% contour. However, this improvement comes at the expense of 
stiffening the equation of state which appears in contradiction with NICER constraints on the radius of a 
$1.4\,{\rm M}_{\odot}$ neutron star\,\cite{Miller:2019cac,Riley:2019yda}. Ultimately, we must conclude that at present 
no single theoretical framework can account simultaneously for the properties of finite nuclei and neutron stars.

\subsection{Coherent elastic neutrino nucleus scattering}
\label{sec:CEvNS}

Shortly after the discovery of weak neutral currents in 1973, coherent elastic neutrino-nucleus scattering (\Cevens\!) 
was proposed as a reaction with favorable cross sections that could impact a variety of astrophysical 
phenomena\,\cite{Freedman:1973yd}. Since then, \Cevens has been recognized as a promising tool for the 
determination of neutron densities and supernovae 
detection\,\cite{Horowitz:2003cz,Scholberg:2005qs,Scholberg:2012id,Patton:2012jr,Patton:2013nwa,Cadeddu:2017etk}. 
The differential cross for a neutrino scattering elastically from a spinless target in the laboratory frame is given 
by\,\cite{Scholberg:2005qs}

\begin{equation}
   \left(\frac{d\sigma}{dT}\right) = \frac{G_{\!F}^{2}}{8\pi} M
   \left[2 - 2 \frac{T}{E} - \frac{MT}{E^{2}} \right]Q_{\rm wk}^{2}F_{\rm wk}^{2}(Q^{2}),
\label{CEvNS}
\end{equation}

where $E$ is the incident neutrino energy, $M$ is the mass of the target nucleus, $T$ the kinetic energy of the recoiling
nucleus, and $Q^{2}\!=\!2MT$. The reaction is ``favorable" because at forward angles the neutrino scatters coherently
from the entire nucleus with a cross section that is proportional to the \emph{square} of the weak charge of the nucleus
$Q^{2}_{\rm wk}\!=\!(NQ_{\rm wk}^{\,n}\!+\!ZQ_{\rm wk}^{\,p})^{2}\!\approx\!N^{2}$. In analogy to Eq.(\ref{dsdOmega}), the 
\Cevens cross section is proportional to the square of the weak charge and probes the distribution of weak charge as 
a function of the momentum transfer. However, unlike elastic electron scattering, the scattered particle is 
undetectable. Hence, rather than detecting the outgoing neutrino one must measure the very low kinetic energy of the 
recoiling nucleus, a heroic feat that took four decades for its experimental realization\,\cite{Akimov:2017ade}. As an
illustration, the maximum kinetic energy of the recoiling nucleus is of the order of $T_{\rm max}\!\simeq\!2E^{2}/M$, 
which for neutrinos produced from pion decay at rest ($E\!\simeq\!30\,{\rm MeV}$) requires detection of nuclear recoils 
with only tens of keVs. 

Besides its impact on a variety of astrophysical phenomena\,\cite{Freedman:1973yd}, \Cevens provides a portal to new
physics. The nuclear weak charge $Q_{\rm wk}$ depends on the weak mixing angle , so a precise measurement of the 
cross section could potentially uncover new physics. However, the required precision is hindered by 
nuclear-structure effects contained in the weak form factor $F_{\rm wk}$, which at low momentum transfers is dominated 
by the weak radius. Although CREX and PREX do not provide direct information on the weak radius of the noble gases 
being used as active targets for the detection of neutrinos as well as dark matter particles, one expects some correlation 
between the structure of neighboring nuclei, such as the weak form factor of ${}^{48}$Ca and the weak radius of ${}^{40}$Ar. 
A preliminary analysis conducted in the pre PREX-2--CREX era suggests a strong correlation between the corresponding 
neutron skin thickness of ${}^{48}$Ca and ${}^{40}$Ar\,\cite{Yang:2019pbx}. A more comprehensive analysis in the post 
PREX-2--CREX era is in progress. We note in closing that a good understanding of \Cevens is essential to assess the 
limitations of dark matter searches, as the so-called ``neutrino floor''---recently renamed the ``neutrino 
fog''---provides an irreducible background that imposes a severe penalty on dark-matter 
detection\,\cite{AristizabalSierra:2019zmy}.
\section{PREX/CREX experimental apparatus}

There have been a number of measurements of nuclear weak form factors using PVES, including 
auxiliary measurements for the Qweak experiment, and dedicated measurements in the PREX/PREX-2 
and CREX campaigns\,\cite{ QweakAl27, PREX2, CREX}.  In this review we focus on the PREX-2 measurement, which has the strongest
impact on the structure of neutron stars, although the experimental methods for CREX are similar.  The experiments ran in Hall A of Jefferson 
Lab (JLab) located in Newport News, VA and use longitudinally polarized electrons that are incident on an isotopically pure solid target. 
Electrons scattered into the Hall A high resolution spectrometers are steered onto the detectors by a set of magnets. The detectors consist 
of thick and thin quartz detectors placed to capture the elastically scattered electrons while minimizing the contribution from inelastic 
processes. Although a more detailed explanation of the PREX and CREX experimental configurations may be found in Ref.\,\cite{PREXlong}, 
a brief summary will be provided here.

\subsection{Parity-violating asymmetry in polarized electron scattering}
\begin{figure}[h]
\includegraphics[width=4in]{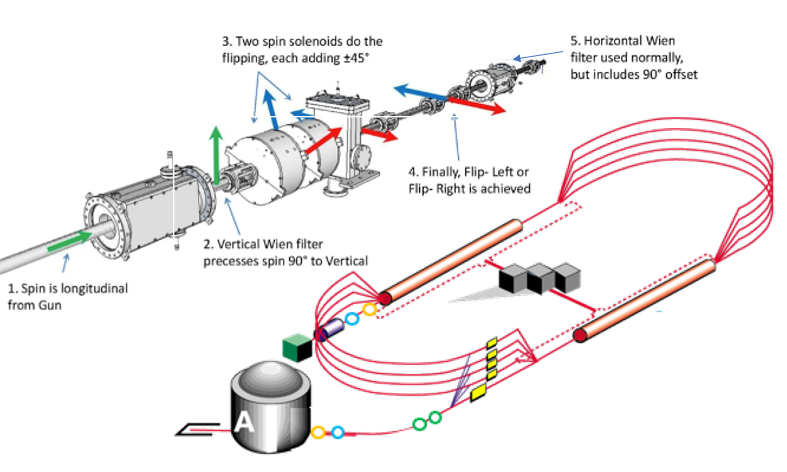}
\caption{The lower right hand side of this figure provides a schematic view of the accelerator. The injector is represented by the green cube 
at the entrance to the north linac, the arcs then bend the electrons into the south linac for a total of six times to reach a maximum energy of 
11 GeV in Hall A.  The upper left hand side of the figure represents a schematic view of the spin manipulator which is located in the injector 
and allows us to change the direction of the spin of the beam to account for spin precession in the accelerator as well as to set examine the
transverse polarization to quantify systematic uncertainties.}
\label{fig:acc}
\end{figure}

Nuclear sizes have long been determined by measuring the charge form factor using parity-conserving elastic electron scattering. These 
experiments have painted the most accurate picture of the distribution of electric charge in atomic nuclei. In order to measure the  
corresponding weak-charge form factor which is largely determined by the neutron distribution, one must incorporate the contribution of
the Z$^0$ boson which couples preferentially to the neutrons in the target. This requires us to measure the parity-violating asymmetry, which
results from the interference of two Feynman diagrams, one mediated by the photon and the other one by the Z$^0$ boson. The ``raw'' or 
measured asymmetry, $A_{\rm meas}$ is the difference in the detector yields between scattering where the spin and momentum of the incident 
electron are parallel (right-handed) and anti-parallel (left-handed) over the sum. That is,
\begin{equation}
A_{meas} = \frac{Y_+-Y_-}{Y_++Y_-}
\end{equation}
Given that the asymmetry involves ratios of cross sections eliminates all corrections that come in as a multiplicative factor. Much 
of the experimental setup is designed to reduce and/or measure beam properties that can introduce a false (non-parity-violating) asymmetry 
into the measured asymmetry. The whole accelerator---which consists of a racetrack with two linear accelerators and arcs at the ends and 
exits into the experimental halls---is depicted in Fig.\ref{fig:acc}. 
As illustrated in the figure, one monitors the degree of polarization of the beam in the injector, which comes in as a dilution factor of the 
measured asymmetry, as well as beam properties such as charge, position, angle, and energy.  In addition, one must correct for the 
asymmetries of background processes (which dilute the $A_{\rm PV}$) and for non-linearities in the detectors.  

The longitudinally polarized electron beam is generated with a circularly polarized laser incident on a strained GaAs photocathode.  The 
resulting electrons are longitudinally polarized, with either positive or negative helicity, depending on the direction of the circular polarization 
of the laser. The direction of the spin can be manipulated with a set of solenoids and Wien filters, and must be directed so that the electrons 
are fully longitudinal when they hit the target, after the spin precession that occurs in the magnets located in the accelerator arcs.  The 
polarization of the laser can be flipped rapidly ($\sim$~2kHz) so that asymmetries can be measured in randomly generated quartets ($+--+$ or $-++-$, 
rather than pairs) on a timescale where the beam properties are consistent in the ``helicity windows'' of the quartet.  The laser polarization is adjusted 
with the use of a Pockels cell, which ideally would not change the position or angle of the laser, but in reality it does. This can introduce, in 
particular, charge asymmetries as the laser in the two polarization states is hitting parts of the photocathode with different quantum efficiencies.  
The charge asymmetry is measured with beam charge monitors throughout the accelerator and is minimized with a feedback system that can 
adjust the position of the laser on the photocathode.  The beam positions, angles and energies are also monitored throughout the accelerator. 

\subsection{Special equipment for PREX}
The PREX and CREX experiments used the Hall A High Resolution Spectrometers (HRSs).  
\begin{figure}[h]
\includegraphics[width=3.5in]{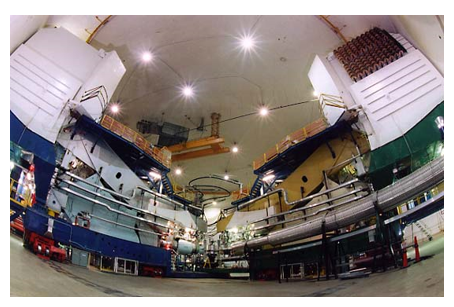}
\caption{Picture of the two spectrometer arms in Hall A, from downstream beam right.  The left HRS is on the right, behind the beamline 
(the grey tube) at a low angle (close to the beamline).  The right HRS is on the left of this picture, at a very large angle relative to the beamline.  
The detector huts are the large, white boxes at the end of each of the HRSs.}
\label{fig:hrs}
\end{figure} 
These consist of two spectrometer ``arms'', left and right (L and R), with limited acceptance in both the horizontal plane (scattering angle) and out of plane 
(azimuthal angle). The spectrometers can rotate around the center of the hall, but their smallest angle relative to the downstream beamline is limited to 
$\geq\!12.5^{\circ}$. Given that the figure-of-merit for PREX is optimized at $\sim$$5^\circ$, the scattering chamber---which usually sits above the pivot 
of the HRSs---was moved upstream and a custom septum magnet was installed just upstream of the HRS to pre-bend the scattered electrons into the HRS.  
A precision collimator in the LHRS, with an exact mirror-image collimator in the RHRS, defines the angular acceptance of each of the arms and ensures that 
they are the same between the two spectrometer arms.  Comparing the asymmetries measured in the LHRS and RHRS allow us to study and/or correct for 
a number of systematic uncertainties, including a parity-conserving asymmetry that depends on the azimuthal angle.

The lead ($^{208}$Pb) target used in PREX was sandwiched between diamond plates and held in place mechanically in a copper frame. The diamond 
backing mitigates the melting of the target through heat transfer to the copper frame. The beam is also rastered on the target to spread the area with the 
heat load.  For the PREX-2 run, the thickest diamond backing tested during PREX was chosen and a ladder with ten targets was prepared.  In order to 
correct for the dilution from the diamond backing, special carbon runs were taken.  
\begin{table}[h!]
\caption{Corrections and systematic uncertainties to the measured PREX-2 parity-violating asymmetry of
              $A^{\rm meas}_{PV}\!=\!550\!\pm\!16 (stat)\!\pm\!8(sys)\,{\rm ppb}$\,\cite{PREX2}.}
\label{tbl:sys_unc}
\begin{center}
\begin{tabular}{ l r r }
\hline
 Correction & Absolute [ppb] & Relative [\%] \\ \hline
Beam asymmetry            & 60.4 $\pm$ 3.0 &  11.0 $\pm$ 0.5 \\
Charge correction         & 20.7 $\pm$ 0.2 &  3.8 $\pm$ 0.0 \\
Beam polarization         & 56.8 $\pm$ 5.2 &  10.3 $\pm$ 1.0 \\
Target diamond foils      &  0.7 $\pm$ 1.4 &  0.1 $\pm$ 0.3 \\
Spectrometer rescattering &  0.0 $\pm$ 0.1 &  0.0 $\pm$ 0.0 \\
Inelastic contributions   &  0.0 $\pm$ 0.1 &  0.0 $\pm$ 0.0 \\
Transverse asymmetry      &  0.0 $\pm$ 0.3 &  0.0 $\pm$ 0.1 \\
Detector nonlinearity     &  0.0 $\pm$ 2.7 &  0.0 $\pm$ 0.5 \\
Angle determination       &  0.0 $\pm$ 3.5 &  0.0 $\pm$ 0.6 \\
Acceptance function       &  0.0 $\pm$ 2.9 &  0.0 $\pm$ 0.5 \\
Total correction          & 17.7 $\pm$ 8.2 &  3.2 $\pm$ 1.5 \\ \hline  
\end{tabular}
\end{center}
\end{table}
The detectors consisted of a set of thick and thin quartz pieces, and a set of GEM trackers for studying the acceptance. Each HRS detector hut hosted 
a set of these detectors.  The detector positions were set carefully in the detector plane to minimize the contribution from inelastic scattering, and the 
amount of rescattering in the spectrometer was studied in special runs. The amount of non-linearity was studied in bench tests. 

We close this section by summarizing in Table\,\ref{tbl:sys_unc} the size of the corrections made to the measured asymmetry and the uncertainties for 
each of the contributions to the systematic uncertainty.  The resulting parity-violating asymmetry is 
$A^{\rm meas}_{PV}\!=\!550\!\pm\!16 (stat)\!\pm\!8(sys)\,{\rm ppb}$\,\cite{PREX2}.  Unlike hadronic experiments that are plagued by model uncertainties,
PREX is a statistically limited experiment.

\newcommand{\APV}{A_{\rm PV}}
\newcommand{\pb}{^{208}{\rm Pb}}
\newcommand{\eca}{^{48}{\rm Ca}}
\newcommand{\Fwk}{F_{\rm wk}}
\newcommand{\Fch}{F_{\rm ch}}
\newcommand{\RW}{R_{\rm W}}
\newcommand{\Rch}{R_{\rm ch}}

\section{Analysis of PREX/CREX data to extract the form factors}
In this section we describe how to extract observables sensitive to the dynamics of neutron rich matter from the $\APV$ measurement 
described in the previous section. In particular, it is important to note that the determination of the weak form factor extracted from $\APV$ 
is as model independent as the corresponding charge form factor extracted from parity-conserving experiments. In both cases Coulomb 
distortions must be taken into account, but these are well known\cite{RocaMaza:2011pm,Horowitz:1998vv,Koshchii:2020qkr}. Other 
quantities, such as the neutron skin, can then be related to the weak form factor, althose these are slightly model dependent. 

\subsection{Calculation of form factors}
To connect the parity-violating asymmetry to the neutron skin thickness one must start by computing various ground state properties.
Given that both $\pb$ and $\eca$ are doubly-magic nuclei, their properties may be computed by solving the mean-field equations in 
the spherical limit\,\cite{Horowitz:1981}. Once the self-consistent solution is obtained, one can extract point proton and neutron densities. 
However, to compare against experiment one must take into account the finite size of the individual nucleons, as $\APV$ depends on both 
the charge and weak form factors. After incorporating single nucleon electromagnetic and weak form factors, the corresponding 
nuclear charge and weak form factors are given by the following expressions\,\cite{Horowitz:2012tj,Horowitz:2012we}:
\begin{eqnarray}
   && Z\Fch(q) = \sum_{i=p,n}\Big(G_E^i(q)F_V^i(q)+\Big(\frac{G_M^i(q)-G_E^i(q)}{1+\tau}\Big) \Big[\tau F_V^i(q)+\frac{q}{2m}F_T^i(q)\Big]\\
   && Q_{\rm{wk}}\Fwk(q) = \sum_{i=p,n}\Big(\widetilde{G}_E^i(q)F_V^i(q)+\Big(\frac{\widetilde{G}_M^i(q)-\widetilde{G}_E^i(q)}{1+\tau}\Big)\Big[\tau F_V^i(q)+\frac{q}{2m}F_T^i(q)\Big],
\end{eqnarray}
where $\tau\!=\!q^2/4m^2$, the sum is over protons and neutrons, $G_E$ and $G_M$ are single-nucleon electric and magnetic Sachs form factors, 
respectively, and the tilde indicates the corresponding weak form factors. In turn, $F_V$ and $F_T$ are (point-nucleon) vector and tensor form factors. 
Nuclear charge and weak-charge densities may now be calculated from the corresponding form factors via a Fourier transform. That is,

\begin{eqnarray}
    \rho_{\mathrm{ch}}(r) = \frac{Z}{2\pi^{2}r}\int_0^\infty q\sin(qr)\Fch(q)dq, \\
    \rho_{\mathrm{wk}}(r) = \frac{Q_{\mathrm{wk}}}{2\pi^{2}r}\int_0^\infty q\sin(qr)\Fwk(q)dq.
\end{eqnarray}
Note that the charge density $\rho_{\mathrm{ch}}(r)$ integrates to the total electric charge $Z$, whereas
the weak-charge density $\rho_{\mathrm{wk}}(r)$ integrates to the total weak charge $Q_{\mathrm{wk}}$.

The plane-wave expression for the parity-violating asymmetry given in Eq.(\ref{APV}) must be modified by including Coulomb distortions.
Electrons scattering off a spherical nucleus satisfy the following Dirac equation:
\begin{eqnarray}
    \Big(\bm{\alpha}\cdot\bm{p}+\beta m_e+\hat{V}(r)\Big)\Psi = E\Psi,
\end{eqnarray}
where $\bm{\alpha}$ and $\beta$ are Dirac matrices and the distorting potential $\hat{V}(r)$ contains both vector and axial-vector components:
\begin{eqnarray}
    \hat{V}(r)=V(r)+\gamma^5 A(r).
\end{eqnarray}
Here $V(r)$ is the standard Coulomb potential satisfying Poisson's equation with the nuclear charge density as its source and $A(r)$ is the 
axial-vector potential given by
\begin{eqnarray}
    A(r)=\frac{G_F}{2^{3/2}}\rho_{\rm wk}(r).
\end{eqnarray} 
To ensure that the model behavior only enters in the weak sector, the nuclear charge density that generates the Coulomb potential is obtained 
from the experimental charge density, extracted from parity-conserving elastic electron scattering\,\cite{deVries:1987}. For $(+/-)$ helicity states, 
the Dirac wavefunction becomes $\Psi_\pm = \frac{1}{2}(1\pm\gamma^5)\Psi$ and the potential becomes
\begin{eqnarray}
    \hat{V}(r) = V(r)\pm A(r).
\end{eqnarray}
In this manner, the positive helicity states scatter from the $V_{+}\!=\!V\!+\!A$ potential, whereas the negative helicity states scatter from the 
$V_{-}\!=\!V\!-\!A$ potential. The difference in the cross sections relative to the sum, yields the parity violating asymmetry as in Eq.(\ref{APV}),
but now with Coulomb distortions taken properly into account.

\subsection{Connecting the model to the measured $\APV$}
Once the model calculations of have been completed, one must convolve the calculated parity-violating asymmetry $\APV^{\rm model}$ with 
the acceptance of the spectrometers in order to compare against the experimental results. That is,
\begin{eqnarray}
    \langle\APV\rangle = \frac{\int d\theta\APV^{\rm model}(\theta)\sin\theta \displaystyle{\frac{d\sigma}{d\Omega}}\epsilon(\theta)}
    {\int d\theta\sin\theta\displaystyle{\frac{d\sigma}{d\Omega}}\epsilon(\theta)},
\end{eqnarray}
where $d\sigma/d\Omega$ is the differential cross section and $\epsilon(\theta)$ is the acceptance function, defined as the relative probability 
for an elastically scattered electron to arrive at the detector. Therefore, each model prediction of the weak-charge density $\rho_{\rm wk}(r)$
generates a unique value for $\APV$ which is then used to extract the nuclear observable and then compare against experiment.
\begin{figure}
    \centering
    \includegraphics[width=4in]{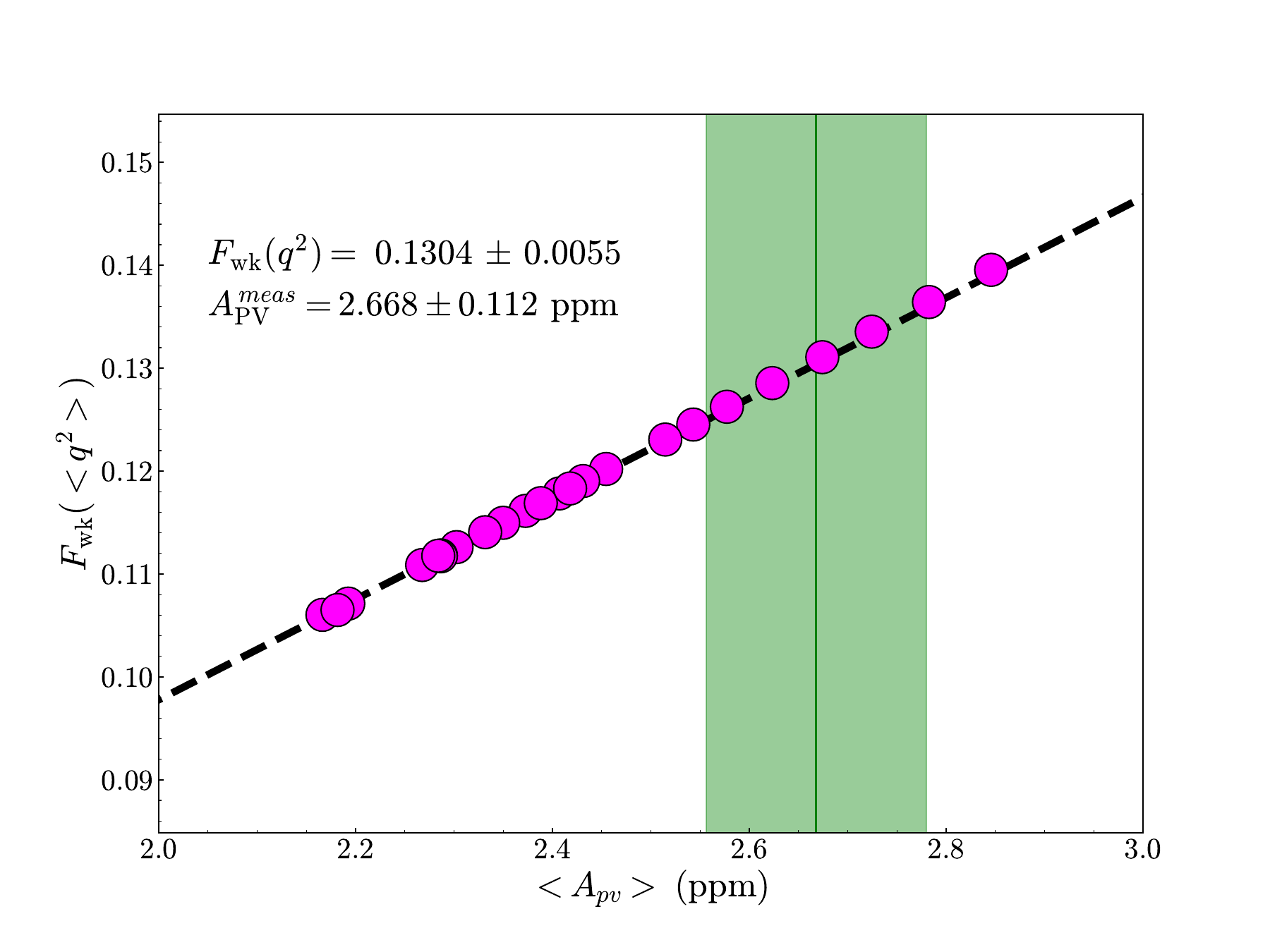}
    \caption{Weak form factor of $\eca$ at the experimental $q^2$ vs. model calculation of $\langle\APV\rangle$. Here the magenta circles are 
    the predictions of several density functionals, the black line is the least-squares fit to the functionals, and the green band represents the 
    measured value of $\APV$ from CREX.}
    \label{fig:FWcrex}
\end{figure}

The cleanest extracted observable from the measured $\APV$ is the weak form factor evaluated at the average $q^2$ of the experiment. As
alluded earlier and shown in Fig.~\ref{fig:FWcrex} for CREX, this is because only the well-known Coulomb distortions affect the plane-wave
expression given in Eq.(\ref{APV}). Given that the weak form factor is model independent, it is quoted without model 
error\,\cite{Adhikari:2021phr,Adhikari:2022kgg}. Although not as strong, the correlation between $\APV$ and the weak radius may be calculated 
via the weak form factor, taking into account the point nucleon densities. By doing so, the neutron skin thickness of ${}^{208}$Pb was determined 
to be\,\cite{PREX2}
\begin{equation}
   R_n - R_p = 0.278 \pm 0.078 ({\rm exp}) \pm 0.012 ({\rm theo})\,{\rm fm},
\end{equation}
where now uncertainties in the surface thickness of ${}^{208}$Pb induce a small model (or theoretical) error.

\section{Future measurements of neutron densities}
In the future, both laboratory experiments and astrophysical observations will further our understanding of the
dynamics of neutron rich matter.  In this section we discuss the Mainz Radius Experiment (MREX) and 
expectations for continued and improved X-ray observations and gravitational wave detections. 

\subsection{The Mainz Radius EXperiment MREX}

\begin{figure}[h]
\includegraphics[width=5in]{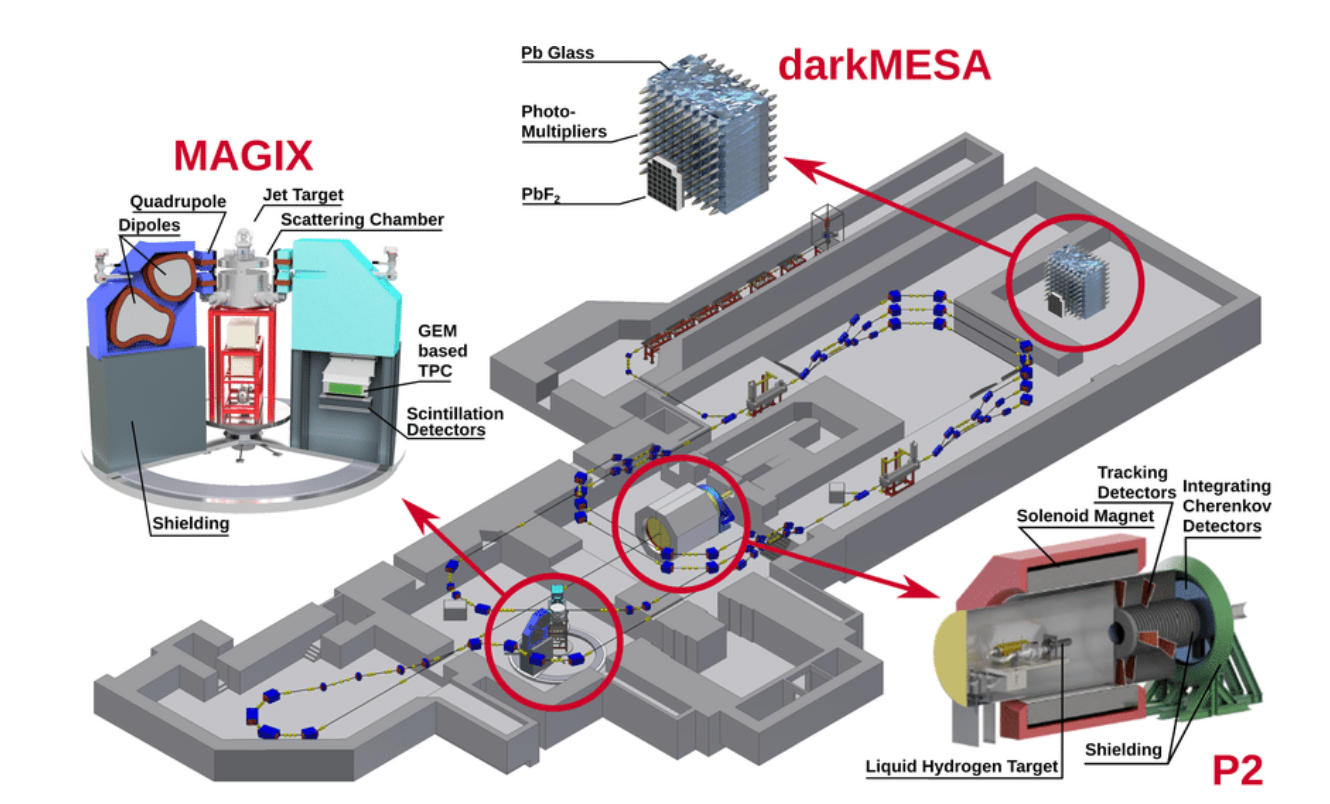}
\caption{Layout of the new MESA facility with the experiments MAGIX, P2, and darkMESA\,\cite{MESANupecc}. 
MREX will use the P2 experimental setup shown in the bottom right corner of the figure. The liquid hydrogen target 
shown will be used for the measurement of the weak mixing angle\,\cite{Becker:2018} and will then be replaced by a 
${}^{208}$Pb target to determine its weak form factor at the same momentum transfer as PREX-2.}
\label{fig:MESA}
\end{figure}

To perform experiments at the intensity and precision frontier, a new accelerator is currently under construction in 
Mainz.  The layout of the Mainz Energy-recovering Superconducting Accelerator (MESA)\,\cite{MESANupecc} is 
shown in Fig.\ref{fig:MESA} together with the some of the planned experiments, such as  MAGIX, P2, and darkMESA. 

The low energy (${\rm E_{max}}\!=\!155\,{\rm MeV}$), high intensity (${\rm I}\!=\!150\,\mu A$), and high-polarization 
(${\rm P}\!\geq\!85\%$) electron beam that MESA will provide offers ideal conditions for future high-precision PVES 
experiments to further constrain the EOS of neutron rich matter, primarily the slope of the symmetry energy $L$. 
The aim of MREX is to determine the neutron skin thickness of ${}^{208}$Pb to better than 0.03\,fm, about a factor 
of two improvement over the precision of PREX-2; see Fig.\ref{fig:WkSkinCompare}. As illustrated in Fig.\ref{fig:MESA},
MREX will use the P2 spectrometer/detector setup\,\cite{Becker:2018} to measure the elastically scattered electrons 
off ${}^{208}$Pb. The measurement will benefit from the full azimuthal coverage of the fused-silica Cherenkov detectors 
due to the solenoidal design of the spectrometer. This will allow for a significant reduction of the statistical uncertainty 
in the determination of the weak form factor and the extracted neutron skin thickness, which is currently the dominating 
contribution to the total uncertainty. 

Preliminary rate estimates suggest that assuming a systematic uncertainty of 1\% in the determination of the parity 
violating asymmetry, a $\pm\,0.03\,{\rm fm}$ determination of the neutron skin thickness is feasible within 1500 hours
of running time. Additional tracking simulations as well as more detailed simulations, including radiative corrections 
and background contributions, are ongoing. In an effort to further constrain the neutron density of ${}^{208}$Pb, we
will determine for the first time its surface thickness by measuring the parity-violating asymmetry at the existing MAMI 
accelerator. Recall that for a heavy nucleus like ${}^{208}$Pb, a two-parameter symmetrized Fermi function accurately 
describes the entire density profile\,\cite{Horowitz:2020evx}; see Eq.(\ref{RhoSF}). Given the current model uncertainty 
of 25\%, the precision in both the PREX and MESA neutron skin determinations are limited by this uncertainty. Thus,
an independent 10\% measurement of the surface thickness will reduce this contribution to a level below the error 
budget of the  $A_{\rm PV}$ measurement\,\cite{Reed:2020}. The kinematics of the experiment is chosen to 
maximize the figure of merit, which in this case---in addition to the sensitivity of $A_{\mathrm{PV}}$ to the weak-charge 
radius---must be suppressed relative to the sensitivity to the surface thickness for the given beam energy and scattering 
angles. Preliminary rate estimates shows that this consideration is fulfilled for two configurations of the spectrometers 
and the beam energies of MAMI. Both possible scenarios are currently under investigation.

\subsection{Future Astrophysical constraints to the equation of state}
\label{Subsection.Nicerfuture}

\begin{figure}[h]
\includegraphics[width=2.75in]{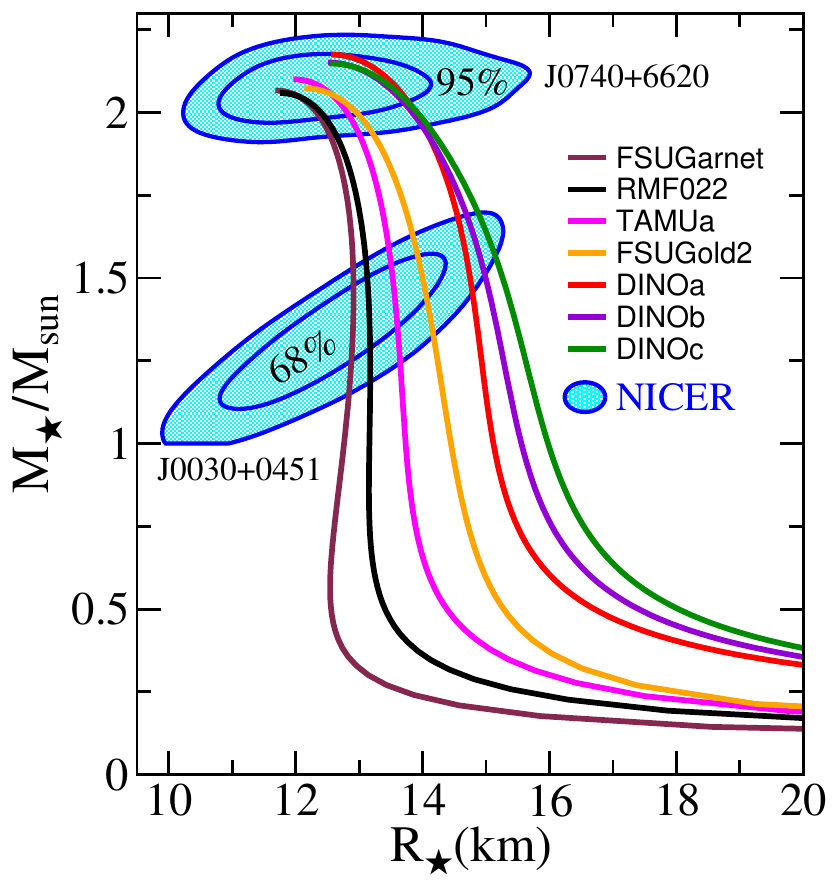}
    \caption{Mass-radius relation for neutron stars. Along with 
    the three DINO models, predictions are also shown for a few 
    covariant energy density functionals. The various ellipses 
    display 68\% and 95\% confidence intervals extracted
    from NICER observations of the two pulsars 
    PSR J0030+0451\,\cite{Riley:2019yda} and 
    PSR J0740+6620\,\cite{Riley:2021pdl}.}
    \label{fig:NS_mr}
\end{figure}

The mass-radius relation, which is in a one-to-one correspondence with the equation of state, is often regarded
as the holy grail of neutron-star structure. We display in Fig.\ref{fig:NS_mr} the mass-radius relation as predicted 
by the recently calibrated DINO models as well as other covariant density functionals that do not include a coupling 
to the $\delta$-meson\,\cite{Reed:2023cap}. The figure also displays simultaneous mass-radius determinations by 
the NICER mission for the two pulsars PSR J0030+045\,\cite{Miller:2019cac,Riley:2019yda} and 
PSR J0740+6620\,\cite{Miller:2021qha,Riley:2021pdl}. With radii of about 15\,km, the predictions from all three 
DINO models are inconsistent at the 68\% level with the NICER data for the low mass star\,\cite{Riley:2019yda}. 
Thus, we conclude that while the DINO models provide a plausible solution to the PREX-2--CREX dilemma, additional 
modifications to the isovector sector must be included in order to reduce its predictions for stellar radii. 

Future X-ray observations of neutron stars will be performed with existing instruments such as NICER and by new missions.  
The high timing and energy resolution of NICER 
has greatly facilitated rationally resolved pulse 
profile measurements that infer the curvature of space.  The European Space Agency Athena mission (scheduled to 
launch in the 2030s) will have a large aperture X-ray telescope and be able to observe fainter sources\,\cite{nandra2013hot}.    
Future X-ray measurements may reduce statistical errors and will provide simultaneous mass-radius results for additional 
neutron stars. X-ray pulse profile measurements may still contain systematic errors due to the unavoidable
modeling of the radiation from pulsar hot spots.  

Gravitational-wave (GW) astronomy has opened a new window into the universe. What sets gravitational waves 
apart is their ability to unveil the inner workings of neutron stars in a unique manner 
complementary to electromagnetic waves. The next-generation GW observatories, such as 
the United States' Cosmic Explorer and Europe's Einstein Telescope, promise unparalleled sensitivity. These 
new observatories will have the capability to detect binary neutron star mergers throughout the entire 
observable universe. Third-generation (3G) gravitational wave observatories will detect countless neutron 
star mergers, enabling the precise determination of stellar radii with a remarkable precision of 100 
meters. Such unprecedented measurements of neutron star radii are poised not only to constrain the elusive 
nuclear equation of state but also to shed light on the composition of their interiors. The 3G era will 
pave the way for mapping the EOS of hot, dense, neutron-rich matter, offering profound new insights into 
the uncharted landscape of Quantum Chromodynamics.

\section{CONCLUSION} 
Neutron stars and atomic nuclei offer fundamental insights into the dynamics of neutron rich matter. Although differing in size by 
18 orders of magnitude, they are both made of the same constituents experiencing the same interactions. The historic detection
of gravitational waves from the binary merger of neutron stars has opened a new window into the universe. Combined with more
precise electromagnetic observations and laboratory experiments, such as PREX-2, CREX, and the future MREX, this new window
will allow us to expand our understanding of neutron rich matter as never before. The dawn of the multi-messenger era has arrived 
and with it the golden era of neutron star physics. These are indeed exhilarating times for nuclear astrophysics.

\section*{DISCLOSURE STATEMENT}
The authors are not aware of any affiliations, memberships, funding, or financial holdings that might be perceived as affecting the objectivity of this review. 

\section*{ACKNOWLEDGMENTS}
We would like to thank the PREX-2/CREX and P2 collaborations. We would also like to thank Anna Watts for providing data for mass-radius contours .   

This material is based upon work which was supported by the German Research Foundation (DFG) via the individual grant Nr. 454637981) and the PRISMA$^+$ Cluster of Excellence, the U.S. Department of Energy Office of Science, Office of Nuclear Physics under Awards DE-FG02-87ER40365 (Indiana University) DE-FG02-92ER40750 (Florida State University), National Science Foundation grant PHY-2116686, Office of Nuclear Physics Contract No. DE-AC05-06OR23177, National Nuclear Security Administration of U.S. Department of Energy (Contract No. 89233218CNA000001), the Laboratory Directed Research and Development program of Los Alamos National Laboratory under project number XXW700, and National Science and Engineering Research Council (NSERC) grant number 322026 (University of Manitoba).

\bibliographystyle{ieeetr}
\bibliography{refs}

\begin{thebibliography}{100}

\bibitem{doi:10.1126/science.1090720}
J.~M. Lattimer and M.~Prakash, ``The physics of neutron stars,'' {\em Science},
  vol.~304, no.~5670, pp.~536--542, 2004.

\bibitem{doi:10.1146/annurev-nucl-102419-124827}
J.~Lattimer, ``Neutron stars and the nuclear matter equation of state,'' {\em
  Annual Review of Nuclear and Particle Science}, vol.~71, no.~1, pp.~433--464,
  2021.

\bibitem{JBell1968}
A.~Hewish, S.~J. Bell, J.~D.~H. Pilkington, P.~F. Scott, and R.~A. Collins,
  ``Observation of a rapidly pulsating radio source,'' {\em Nature}, vol.~217,
  no.~5130, pp.~709--713, 1968.

\bibitem{peeling}
A.~Lam {\em et~al.}, ``Peeling apart a neutron star.''
  \url{https://astrobites.org/2014/08/11/peeling-apart-a-neutron-star/}, 2014.
\newblock [Online; accessed September 1, 2023].

\bibitem{PhysRev.55.374}
J.~R. Oppenheimer and G.~M. Volkoff, ``On massive neutron cores,'' {\em Phys.
  Rev.}, vol.~55, pp.~374--381, Feb 1939.

\bibitem{Vidana:2000mx}
I.~Vidana, A.~Polls, A.~Ramos, L.~Engvik, and M.~Hjorth-Jensen, ``{Hyperon
  effects on the properties of beta stable neutron star matter},'' {\em Nucl.
  Phys. A}, vol.~691, pp.~443--446, 2001.

\bibitem{Oertel:2016xsn}
M.~Oertel, F.~Gulminelli, C.~Provid\^encia, and A.~R. Raduta, ``{Hyperons in
  neutron stars and supernova cores},'' {\em Eur. Phys. J. A}, vol.~52, no.~3,
  p.~50, 2016.

\bibitem{Oertel:2016bki}
M.~Oertel, M.~Hempel, T.~Kl\"ahn, and S.~Typel, ``{Equations of state for
  supernovae and compact stars},'' {\em Rev. Mod. Phys.}, vol.~89, no.~1,
  p.~015007, 2017.

\bibitem{Demorest:2010bx}
P.~Demorest, T.~Pennucci, S.~Ransom, M.~Roberts, and J.~Hessels, ``{Shapiro
  delay measurement of a two solar mass neutron star},'' {\em Nature},
  vol.~467, p.~1081, 2010.

\bibitem{Antoniadis:2013pzd}
J.~Antoniadis, P.~C. Freire, N.~Wex, T.~M. Tauris, R.~S. Lynch, {\em et~al.},
  ``{A Massive Pulsar in a Compact Relativistic Binary},'' {\em Science},
  vol.~340, p.~6131, 2013.

\bibitem{Cromartie:2019kug}
H.~T. Cromartie {\em et~al.}, ``{Relativistic Shapiro delay measurements of an
  extremely massive millisecond pulsar},'' {\em Nat. Astron.}, vol.~4, no.~1,
  pp.~72--76, 2019.

\bibitem{Fonseca:2021wxt}
E.~Fonseca {\em et~al.}, ``{Refined Mass and Geometric Measurements of the
  High-mass PSR J0740+6620},'' {\em Astrophys. J. Lett.}, vol.~915, no.~1,
  p.~L12, 2021.

\bibitem{Alford:2007xm}
M.~G. Alford, A.~Schmitt, K.~Rajagopal, and T.~Schafer, ``{Color
  superconductivity in dense quark matter},'' {\em Rev. Mod. Phys.}, vol.~80,
  pp.~1455--1515, 2008.

\bibitem{Brown:2000}
B.~A. Brown, ``Neutron radii in nuclei and the neutron equation of state,''
  {\em Phys. Rev. Lett.}, vol.~85, p.~5296, 2000.

\bibitem{Furnstahl:2001un}
R.~J. Furnstahl, ``Neutron radii in mean-field models,'' {\em Nucl. Phys.},
  vol.~A706, pp.~85--110, 2002.

\bibitem{Centelles:2008vu}
M.~Centelles, X.~Roca-Maza, X.~Vi\~nas, and M.~Warda, ``{Nuclear symmetry
  energy probed by neutron skin thickness of nuclei},'' {\em Phys. Rev. Lett.},
  vol.~102, p.~122502, 2009.

\bibitem{RocaMaza:2011pm}
X.~Roca-Maza, M.~Centelles, X.~Vi\~nas, and M.~Warda, ``{Neutron skin of
  $^{208}Pb$, nuclear symmetry energy, and the parity radius experiment},''
  {\em Phys. Rev. Lett.}, vol.~106, p.~252501, 2011.

\bibitem{Piekarewicz:2008nh}
J.~Piekarewicz and M.~Centelles, ``{Incompressibility of neutron-rich
  matter},'' {\em Phys. Rev.}, vol.~C79, p.~054311, 2009.

\bibitem{Miller:2019cac}
M.~C. Miller {\em et~al.}, ``{PSR J0030+0451 Mass and Radius from NICER Data
  and Implications for the Properties of Neutron Star Matter},'' {\em
  Astrophys. J. Lett.}, vol.~887, p.~L24, 2019.

\bibitem{Riley:2019yda}
T.~E. Riley {\em et~al.}, ``{A NICER View of PSR J0030+0451: Millisecond Pulsar
  Parameter Estimation},'' {\em Astrophys. J. Lett.}, vol.~887, p.~L21, 2019.

\bibitem{Miller:2021qha}
M.~C. Miller {\em et~al.}, ``{The Radius of PSR J0740+6620 from NICER and
  XMM-Newton Data},'' {\em Astrophys. J. Lett.}, vol.~918, no.~2, p.~L28, 2021.

\bibitem{Abbott:PRL2017}
B.~P. Abbott {\em et~al.}, ``{GW170817: Observation of Gravitational Waves from
  a Binary Neutron Star Inspiral},'' {\em Phys. Rev. Lett.}, vol.~119, no.~16,
  p.~161101, 2017.

\bibitem{Drout:2017ijr}
M.~R. Drout {\em et~al.}, ``{Light Curves of the Neutron Star Merger
  GW170817/SSS17a: Implications for R-Process Nucleosynthesis},'' {\em
  Science}, vol.~358, pp.~1570--1574, 2017.

\bibitem{Cowperthwaite:2017dyu}
P.~S. Cowperthwaite {\em et~al.}, ``{The Electromagnetic Counterpart of the
  Binary Neutron Star Merger LIGO/Virgo GW170817. II. UV, Optical, and
  Near-infrared Light Curves and Comparison to Kilonova Models},'' {\em
  Astrophys. J.}, vol.~848, no.~2, p.~L17, 2017.

\bibitem{Chornock:2017sdf}
R.~Chornock {\em et~al.}, ``{The Electromagnetic Counterpart of the Binary
  Neutron Star Merger LIGO/VIRGO GW170817. IV. Detection of Near-infrared
  Signatures of r-process Nucleosynthesis with Gemini-South},'' {\em Astrophys.
  J.}, vol.~848, no.~2, p.~L19, 2017.

\bibitem{Nicholl:2017ahq}
M.~Nicholl {\em et~al.}, ``{The Electromagnetic Counterpart of the Binary
  Neutron Star Merger LIGO/VIRGO GW170817. III. Optical and UV Spectra of a
  Blue Kilonova From Fast Polar Ejecta},'' {\em Astrophys. J.}, vol.~848,
  no.~2, p.~L18, 2017.

\bibitem{Bauswein:2017vtn}
A.~Bauswein, O.~Just, H.-T. Janka, and N.~Stergioulas, ``{Neutron-star radius
  constraints from GW170817 and future detections},'' {\em Astrophys. J.},
  vol.~850, no.~2, p.~L34, 2017.

\bibitem{Fattoyev:2017jql}
F.~J. Fattoyev, J.~Piekarewicz, and C.~J. Horowitz, ``{Neutron skins and
  neutron stars in the multi-messenger era},'' {\em Phys. Rev. Lett.},
  vol.~120, no.~17, p.~172702, 2018.

\bibitem{Annala:2017llu}
E.~Annala, T.~Gorda, A.~Kurkela, and A.~Vuorinen, ``{Gravitational-wave
  constraints on the neutron-star-matter Equation of State},'' {\em Phys. Rev.
  Lett.}, vol.~120, no.~17, p.~172703, 2018.

\bibitem{Abbott:2018exr}
B.~P. Abbott {\em et~al.}, ``{GW170817: Measurements of neutron star radii and
  equation of state},'' {\em Phys. Rev. Lett.}, vol.~121, no.~16, p.~161101,
  2018.

\bibitem{Most:2018hfd}
E.~R. Most, L.~R. Weih, L.~Rezzolla, and J.~Schaffner-Bielich, ``{New
  constraints on radii and tidal deformabilities of neutron stars from
  GW170817},'' {\em Phys. Rev. Lett.}, vol.~120, no.~26, p.~261103, 2018.

\bibitem{Tews:2018chv}
I.~Tews, J.~Margueron, and S.~Reddy, ``{Critical examination of constraints on
  the equation of state of dense matter obtained from GW170817},'' {\em Phys.
  Rev.}, vol.~C98, no.~4, p.~045804, 2018.

\bibitem{Malik:2018zcf}
T.~Malik, N.~Alam, M.~Fortin, C.~Providencia, B.~K. Agrawal, T.~K. Jha,
  B.~Kumar, and S.~K. Patra, ``{GW170817: constraining the nuclear matter
  equation of state from the neutron star tidal deformability},'' {\em Phys.
  Rev.}, vol.~C98, no.~3, p.~035804, 2018.

\bibitem{Radice:2017lry}
D.~Radice, A.~Perego, F.~Zappa, and S.~Bernuzzi, ``{GW170817: Joint Constraint
  on the Neutron Star Equation of State from Multimessenger Observations},''
  {\em Astrophys. J. Lett.}, vol.~852, no.~2, p.~L29, 2018.

\bibitem{Radice:2018ozg}
D.~Radice and L.~Dai, ``{Multimessenger Parameter Estimation of GW170817},''
  {\em Eur. Phys. J. A}, vol.~55, no.~4, p.~50, 2019.

\bibitem{Tews:2019cap}
I.~Tews, J.~Margueron, and S.~Reddy, ``{Confronting gravitational-wave
  observations with modern nuclear physics constraints},'' {\em Eur. Phys. J.
  A}, vol.~55, no.~6, p.~97, 2019.

\bibitem{Capano:2019eae}
C.~D. Capano, I.~Tews, S.~M. Brown, B.~Margalit, S.~De, S.~Kumar, D.~A. Brown,
  B.~Krishnan, and S.~Reddy, ``{Stringent constraints on neutron-star radii
  from multimessenger observations and nuclear theory},'' {\em Nature
  Astronomy}, vol.~4, pp.~625--632, 8 2019.

\bibitem{Tsang:2019mlz}
M.~Tsang, W.~Lynch, P.~Danielewicz, and C.~Tsang, ``{Symmetry energy
  constraints from GW170817 and laboratory experiments},'' {\em Phys. Lett. B},
  vol.~795, pp.~533--536, 2019.

\bibitem{Tsang:2020lmb}
C.~Y. Tsang, M.~B. Tsang, P.~Danielewicz, W.~G. Lynch, and F.~J. Fattoyev,
  ``{Impact of the neutron-star deformability on equation of state
  parameters},'' {\em Phys. Rev. C}, vol.~102, no.~4, p.~045808, 2020.

\bibitem{Drischler:2020hwi}
C.~Drischler, R.~Furnstahl, J.~Melendez, and D.~Phillips, ``{How Well Do We
  Know the Neutron-Matter Equation of State at the Densities Inside Neutron
  Stars? A Bayesian Approach with Correlated Uncertainties},'' {\em Phys. Rev.
  Lett.}, vol.~125, no.~20, p.~202702, 2020.

\bibitem{Landry:2020vaw}
P.~Landry, R.~Essick, and K.~Chatziioannou, ``{Nonparametric constraints on
  neutron star matter with existing and upcoming gravitational wave and pulsar
  observations},'' {\em Phys. Rev. D}, vol.~101, no.~12, p.~123007, 2020.

\bibitem{Xie:2020rwg}
W.-J. Xie and B.-A. Li, ``{Bayesian inference of the dense-matter equation of
  state encapsulating a first-order hadron-quark phase transition from
  observables of canonical neutron stars},'' {\em Phys. Rev. C}, vol.~103,
  no.~3, p.~035802, 2021.

\bibitem{Essick:2021kjb}
R.~Essick, I.~Tews, P.~Landry, and A.~Schwenk, ``{Astrophysical Constraints on
  the Symmetry Energy and the Neutron Skin of Pb208 with Minimal Modeling
  Assumptions},'' {\em Phys. Rev. Lett.}, vol.~127, no.~19, p.~192701, 2021.

\bibitem{Chatziioannou:2021tdi}
K.~Chatziioannou, ``{Uncertainty limits on neutron star radius measurements
  with gravitational waves},'' {\em Phys. Rev. D}, vol.~105, no.~8, p.~084021,
  2022.

\bibitem{Reed:2021nqk}
B.~T. Reed, F.~J. Fattoyev, C.~J. Horowitz, and J.~Piekarewicz, ``{Implications
  of PREX-II on the equation of state of neutron-rich matter},'' {\em Phys.
  Rev. Lett.}, vol.~126, no.~17, p.~172503, 2021.

\bibitem{Sammarruca:2022ser}
F.~Sammarruca and R.~Millerson, ``{The Equation of State of Neutron-Rich Matter
  at Fourth Order of Chiral Effective Field Theory and the Radius of a
  Medium-Mass Neutron Star},'' {\em Universe}, vol.~8, no.~2, p.~133, 2022.

\bibitem{Damour:1991yw}
T.~Damour, M.~Soffel, and C.-m. Xu, ``{General relativistic celestial
  mechanics. 2. Translational equations of motion},'' {\em Phys. Rev.},
  vol.~D45, pp.~1017--1044, 1992.

\bibitem{Flanagan:2007ix}
E.~E. Flanagan and T.~Hinderer, ``{Constraining neutron star tidal Love numbers
  with gravitational wave detectors},'' {\em Phys. Rev.}, vol.~D77, p.~021502,
  2008.

\bibitem{Love:1909}
A.~E.~H. Love, ``The yielding of the earth to disturbing forces,'' {\em Proc.
  R. Soc. Lond.}, vol.~A82, pp.~73--88, 1909.

\bibitem{Binnington:2009bb}
T.~Binnington and E.~Poisson, ``{Relativistic theory of tidal Love numbers},''
  {\em Phys. Rev.}, vol.~D80, p.~084018, 2009.

\bibitem{Damour:2012yf}
T.~Damour, A.~Nagar, and L.~Villain, ``{Measurability of the tidal
  polarizability of neutron stars in late-inspiral gravitational-wave
  signals},'' {\em Phys. Rev.}, vol.~D85, p.~123007, 2012.

\bibitem{Hinderer:2007mb}
T.~Hinderer, ``{Tidal Love numbers of neutron stars},'' {\em Astrophys. J.},
  vol.~677, pp.~1216--1220, 2008.

\bibitem{Hinderer:2009ca}
T.~Hinderer, B.~D. Lackey, R.~N. Lang, and J.~S. Read, ``{Tidal deformability
  of neutron stars with realistic equations of state and their gravitational
  wave signatures in binary inspiral},'' {\em Phys. Rev.}, vol.~D81, p.~123016,
  2010.

\bibitem{Damour:2009vw}
T.~Damour and A.~Nagar, ``{Relativistic tidal properties of neutron stars},''
  {\em Phys. Rev.}, vol.~D80, p.~084035, 2009.

\bibitem{Postnikov:2010yn}
S.~Postnikov, M.~Prakash, and J.~M. Lattimer, ``{Tidal Love Numbers of Neutron
  and Self-Bound Quark Stars},'' {\em Phys. Rev.}, vol.~D82, p.~024016, 2010.

\bibitem{Fattoyev:2012uu}
F.~J. Fattoyev, J.~Carvajal, W.~G. Newton, and B.-A. Li, ``{Constraining the
  high-density behavior of the nuclear symmetry energy with the tidal
  polarizability of neutron stars},'' {\em Phys. Rev.}, vol.~C87, no.~1,
  p.~015806, 2013.

\bibitem{Steiner:2014pda}
A.~W. Steiner, S.~Gandolfi, F.~J. Fattoyev, and W.~G. Newton, ``{Using Neutron
  Star Observations to Determine Crust Thicknesses, Moments of Inertia, and
  Tidal Deformabilities},'' {\em Phys. Rev.}, vol.~C91, no.~1, p.~015804, 2015.

\bibitem{Piekarewicz:2018sgy}
J.~Piekarewicz and F.~J. Fattoyev, ``{Impact of the neutron star crust on the
  tidal polarizability},'' {\em Phys. Rev. C}, vol.~99, no.~4, p.~045802, 2019.

\bibitem{Hofstadter:1956qs}
R.~Hofstadter, ``{Electron scattering and nuclear structure},'' {\em Rev. Mod.
  Phys.}, vol.~28, pp.~214--254, 1956.

\bibitem{DeJager:1987qc}
H.~De~Vries, C.~W. De~Jager, and C.~De~Vries, ``{Nuclear charge and
  magnetization density distribution parameters from elastic electron
  scattering},'' {\em Atom. Data Nucl. Data Tabl.}, vol.~36, pp.~495--536,
  1987.

\bibitem{Fricke:1995}
G.~Fricke, C.~Bernhardt, K.~Heilig, L.~A. Schaller, L.~Schellenberg, E.~B.
  Shera, and C.~W. de~Jager, ``Nuclear ground state charge radii from
  electromagnetic interactions,'' {\em Atom. Data and Nucl. Data Tables},
  vol.~60, p.~177, 1995.

\bibitem{Angeli:2013}
I.~Angeli and K.~Marinova, ``Table of experimental nuclear ground state charge
  radii: An update,'' {\em At. Data Nucl. Data Tables}, vol.~99, pp.~69 -- 95,
  2013.

\bibitem{Suda:2009zz}
T.~Suda {\em et~al.}, ``{First Demonstration of Electron Scattering Using a
  Novel Target Developed for Short-Lived Nuclei},'' {\em Phys. Rev. Lett.},
  vol.~102, p.~102501, 2009.

\bibitem{Suda:2017nss}
T.~Suda and H.~Simon, ``{Prospects for electron scattering on unstable, exotic
  nuclei},'' {\em Prog. Part. Nucl. Phys.}, vol.~96, pp.~1--31, 2017.

\bibitem{Aitchinson:1982}
I.~J.~R. Aitchison and A.~J.~G. Hey, {\em Gauge Theories in Particle Physics,
  Second Edition}.
\newblock Taylor \& Francis, 1989.

\bibitem{Walecka:2001}
J.~D. Walecka, {\em Electron scattering for nuclear and nucleon structure}.
\newblock New York: Cambridge University Press, 2001.

\bibitem{Sprung:1997}
D.~W. Sprung and J.~Martorell, ``{The symmetrized Fermi function and its
  transforms},'' {\em J. Phys. A}, vol.~30, pp.~6525--6534, 1997.

\bibitem{Piekarewicz:2016vbn}
J.~Piekarewicz, A.~R. Linero, P.~Giuliani, and E.~Chicken, ``{Power of two:
  Assessing the impact of a second measurement of the weak-charge form factor
  of $^{208}$Pb},'' {\em Phys. Rev.}, vol.~C94, no.~3, p.~034316, 2016.

\bibitem{Amado:1979st}
R.~D. Amado, J.~P. Dedonder, and F.~Lenz, ``{An Explicit Formula for Hadron -
  Nucleus Elastic Scattering in the Eikonal Approximation},'' {\em Phys. Rev.},
  vol.~C21, pp.~647--661, 1980.

\bibitem{Amado:1986pm}
R.~D. Amado, ``{Analytic insights into intermediate-energy hadron nucleus
  scattering},'' {\em Adv. Nucl. Phys.}, vol.~15, pp.~1--42, 1985.

\bibitem{Thiel:2019tkm}
M.~Thiel, C.~Sfienti, J.~Piekarewicz, C.~J. Horowitz, and M.~Vanderhaeghen,
  ``{Neutron skins of atomic nuclei: per aspera ad astra},'' {\em J. Phys.},
  vol.~G46, no.~9, p.~093003, 2019.

\bibitem{Donnelly:1989qs}
T.~Donnelly, J.~Dubach, and I.~Sick, ``{Isospin dependences in parity violating
  electron scattering},'' {\em Nucl. Phys.}, vol.~A503, p.~589, 1989.

\bibitem{Abrahamyan:2012gp}
S.~Abrahamyan, Z.~Ahmed, H.~Albataineh, K.~Aniol, D.~S. Armstrong, {\em
  et~al.}, ``{Measurement of the Neutron Radius of 208Pb Through
  Parity-Violation in Electron Scattering},'' {\em Phys. Rev. Lett.}, vol.~108,
  p.~112502, 2012.

\bibitem{Horowitz:2012tj}
C.~J. Horowitz, Z.~Ahmed, C.~M. Jen, A.~Rakhman, P.~A. Souder, {\em et~al.},
  ``{Weak charge form factor and radius of 208Pb through parity violation in
  electron scattering},'' {\em Phys. Rev.}, vol.~C85, p.~032501, 2012.

\bibitem{Androic:2018kni}
D.~Androic {\em et~al.}, ``{Precision measurement of the weak charge of the
  proton},'' {\em Nature}, vol.~557, no.~7704, pp.~207--211, 2018.

\bibitem{Horowitz:1998vv}
C.~J. Horowitz, ``{Parity violating elastic electron scattering and Coulomb
  distortions},'' {\em Phys. Rev.}, vol.~C57, pp.~3430--3436, 1998.

\bibitem{RocaMaza:2008cg}
X.~Roca-Maza, M.~Centelles, F.~Salvat, and X.~Vinas, ``{Theoretical study of
  elastic electron scattering off stable and exotic nuclei},'' {\em Phys.
  Rev.}, vol.~C78, p.~044332, 2008.

\bibitem{Koshchii:2020qkr}
O.~Koshchii, J.~Erler, M.~Gorchtein, C.~J. Horowitz, J.~Piekarewicz,
  X.~Roca-Maza, C.-Y. Seng, and H.~Spiesberger, ``{Weak charge and weak radius
  of $^{12}$C},'' {\em Phys. Rev. C}, vol.~102, no.~2, p.~022501, 2020.

\bibitem{Hagen:2015yea}
G.~Hagen {\em et~al.}, ``{Neutron and weak-charge distributions of the
  $^{48}$Ca nucleus},'' {\em Nature Phys.}, vol.~12, no.~2, pp.~186--190, 2016.

\bibitem{Mahzoon:2017fsg}
M.~H. Mahzoon, M.~C. Atkinson, R.~J. Charity, and W.~H. Dickhoff, ``{Neutron
  Skin Thickness of $^{48}$Ca from a Nonlocal Dispersive Optical-Model
  Analysis},'' {\em Phys. Rev. Lett.}, vol.~119, no.~22, p.~222503, 2017.

\bibitem{Adhikari:2022kgg}
D.~Adhikari {\em et~al.}, ``{Precision Determination of the Neutral Weak Form
  Factor of Ca48},'' {\em Phys. Rev. Lett.}, vol.~129, no.~4, p.~042501, 2022.

\bibitem{Adhikari:2021phr}
D.~Adhikari {\em et~al.}, ``{Accurate Determination of the Neutron Skin
  Thickness of $^{208}$Pb through Parity-Violation in Electron Scattering},''
  {\em Phys. Rev. Lett.}, vol.~126, no.~17, p.~172502, 2021.

\bibitem{Carriere:2002bx}
J.~Carriere, C.~J. Horowitz, and J.~Piekarewicz, ``Low mass neutron stars and
  the equation of state of dense matter,'' {\em Astrophys. J.}, vol.~593,
  p.~463, 2003.

\bibitem{Horowitz:2012we}
C.~J. Horowitz and J.~Piekarewicz, ``{Impact of spin-orbit currents on the
  electroweak skin of neutron-rich nuclei},'' {\em Phys. Rev.}, vol.~C86,
  p.~045503, 2012.

\bibitem{Piekarewicz:2012pp}
J.~Piekarewicz, B.~Agrawal, G.~Col\`o, W.~Nazarewicz, N.~Paar, {\em et~al.},
  ``{Electric dipole polarizability and the neutron skin},'' {\em Phys. Rev.},
  vol.~C85, p.~041302(R), 2012.

\bibitem{Piekarewicz:2021jte}
J.~Piekarewicz, ``{Implications of PREX-2 on the electric dipole polarizability
  of neutron-rich nuclei},'' {\em Phys. Rev. C}, vol.~104, no.~2, p.~024329,
  2021.

\bibitem{Reed:2023cap}
B.~T. Reed, F.~J. Fattoyev, C.~J. Horowitz, and J.~Piekarewicz, ``{Density
  Dependence of the Symmetry Energy in the Post PREX-CREX Era},'' {\em
  2305.19376}, 2023.

\bibitem{Freedman:1973yd}
D.~Z. Freedman, ``{Coherent Neutrino Nucleus Scattering as a Probe of the Weak
  Neutral Current},'' {\em Phys. Rev.}, vol.~D9, pp.~1389--1392, 1974.

\bibitem{Horowitz:2003cz}
C.~J. Horowitz, K.~J. Coakley, and D.~N. McKinsey, ``{Supernova observation via
  neutrino - nucleus elastic scattering in the CLEAN detector},'' {\em Phys.
  Rev.}, vol.~D68, p.~023005, 2003.

\bibitem{Scholberg:2005qs}
K.~Scholberg, ``{Prospects for measuring coherent neutrino-nucleus elastic
  scattering at a stopped-pion neutrino source},'' {\em Phys. Rev.}, vol.~D73,
  p.~033005, 2006.

\bibitem{Scholberg:2012id}
K.~Scholberg, ``{Supernova Neutrino Detection},'' {\em Ann. Rev. Nucl. Part.
  Sci.}, vol.~62, pp.~81--103, 2012.

\bibitem{Patton:2012jr}
K.~Patton, J.~Engel, G.~C. McLaughlin, and N.~Schunck, ``{Neutrino-nucleus
  coherent scattering as a probe of neutron density distributions},'' {\em
  Phys. Rev.}, vol.~C86, p.~024612, 2012.

\bibitem{Patton:2013nwa}
K.~M. Patton, G.~C. McLaughlin, and K.~Scholberg, ``{Prospects for using
  coherent elastic neutrino-nucleus scattering to measure the nuclear neutron
  form factor},'' {\em Int. J. Mod. Phys.}, vol.~E22, p.~1330013, 2013.

\bibitem{Cadeddu:2017etk}
M.~Cadeddu, C.~Giunti, Y.~F. Li, and Y.~Y. Zhang, ``{Average CsI neutron
  density distribution from COHERENT data},'' {\em Phys. Rev. Lett.}, vol.~120,
  no.~7, p.~072501, 2018.

\bibitem{Akimov:2017ade}
D.~Akimov {\em et~al.}, ``{Observation of Coherent Elastic Neutrino-Nucleus
  Scattering},'' {\em Science}, vol.~357, no.~6356, pp.~1123--1126, 2017.

\bibitem{Yang:2019pbx}
J.~Yang, J.~A. Hernandez, and J.~Piekarewicz, ``{Electroweak probes of ground
  state densities},'' {\em Phys. Rev. C}, vol.~100, no.~5, p.~054301, 2019.

\bibitem{AristizabalSierra:2019zmy}
D.~Aristizabal~Sierra, J.~Liao, and D.~Marfatia, ``{Impact of form factor
  uncertainties on interpretations of coherent elastic neutrino-nucleus
  scattering data},'' {\em 1902.07398}, 2019.

\bibitem{QweakAl27}
D.~Androic {\em et~al.}, ``{Determination of the $^{27}$AI Neutron Distribution
  Radius from a Parity-Violating Electron Scattering Measurement},'' {\em Phys.
  Rev. Lett.}, vol.~128, no.~13, p.~132501, 2022.

\bibitem{PREX2}
D.~Adhikari {\em et~al.}, ``{Accurate Determination of the Neutron Skin
  Thickness of $^{208}$Pb through Parity-Violation in Electron Scattering},''
  {\em Phys. Rev. Lett.}, vol.~126, no.~17, p.~172502, 2021.

\bibitem{CREX}
D.~Adhikari {\em et~al.}, ``{Precision Determination of the Neutral Weak Form
  Factor of Ca48},'' {\em Phys. Rev. Lett.}, vol.~129, no.~4, p.~042501, 2022.

\bibitem{PREXlong}
D.~Adhikari {\em et~al.}, ``{Precision Neutron Skins of $^{208}$Pb and
  $^{48}$Ca from Parity-Violating Electron Scattering},'' {\em in preparation},
  2023.

\bibitem{Horowitz:1981}
C.~J. {Horowitz} and B.~D. {Serot}, ``{Self-consistent hartree description of
  finite nuclei in a relativistic quantum field theory},'' {\em Nuclear
  Physics}, vol.~A 368, pp.~503--528, Oct. 1981.

\bibitem{deVries:1987}
H.~{de Vries}, C.~W. {de Jager}, and C.~{de Vries}, ``{Nuclear
  Charge-Density-Distribution Parameters from Electron Scattering},'' {\em
  Atomic Data and Nuclear Data Tables}, vol.~36, p.~495, Jan. 1987.

\bibitem{MESANupecc}
N.~B. et~al., ``{The MESA Experimental Program: A Laboratory for Precision
  Physics with Electron Scattering at Low Energy”},'' {\em Nucl. Phys. News},
  vol.~31, no.~3, p.~5, 2021.

\bibitem{Becker:2018}
D.~B. et~al., ``{The P2 experiment},'' {\em Eur. Phys. J. A}, vol.~54, no.~11,
  p.~208, 2018.

\bibitem{Horowitz:2020evx}
C.~J. Horowitz, J.~Piekarewicz, and B.~Reed, ``{Insights into nuclear
  saturation density from parity violating electron scattering},'' {\em Phys.
  Rev. C}, vol.~102, no.~4, p.~044321, 2020.

\bibitem{Reed:2020}
B.~Reed, Z.~Jaffe, C.~J. Horowitz, and C.~Sfienti, ``{Measuring the surface
  thickness of the weak charge density of nuclei},'' {\em Phys. Rev. C},
  vol.~102, p.~064308, 2020.

\bibitem{Riley:2021pdl}
T.~E. Riley {\em et~al.}, ``{A NICER View of the Massive Pulsar PSR J0740+6620
  Informed by Radio Timing and XMM-Newton Spectroscopy},'' {\em Astrophys. J.
  Lett.}, vol.~918, no.~2, p.~L27, 2021.

\bibitem{nandra2013hot}
K.~Nandra and et~al., ``The hot and energetic universe: A white paper
  presenting the science theme motivating the athena+ mission,'' {\em
  1306.2307}, 2013.

\end{thebibliography}

\end{document}